\documentclass[10pt,aps,showpacs,nofootinbib,prd,aps,epsf,floats,
               amsmath,amssymb,amsfonts,axodraw,floatfix,graphicx,twocolumn]{revtex4-1}

\usepackage{amsmath, amssymb}
\usepackage{axodraw}
\usepackage{multirow}
\usepackage{paralist}
\newcommand{\mathsym}[1]{{}}

\usepackage{graphicx}
\usepackage{amsmath}
\usepackage{amssymb}
\usepackage{amsmath}
\usepackage{multirow}
\usepackage{paralist}
\usepackage{slashed}
\usepackage{amsfonts}
\usepackage{hyperref} 
\usepackage{graphicx}
\usepackage{amsmath}
\usepackage{amssymb}
\usepackage{mathrsfs}
\newsavebox{\PSLASH}
 \sbox{\PSLASH}{$p$\hspace{-1.8mm}/}
 
\renewcommand{\theequation}{\thesection.\arabic{equation}}
\newcounter{saveeqn}
\newcommand{\add}{\addtocounter{equation}{1}}
\newcommand{\alphaeqn}{\setcounter{saveeqn}{\value{equation}}%
\setcounter{equation}{0}%
\renewcommand{\theequation}{\mbox{\thesection.\arabic{saveeqn}{\alpha{equation}}}}}
\newcommand{\reseteqn}{\setcounter{equation}{\value{saveeqn}}%
\renewcommand{\theequation}{\thesection.\arabic{equation}}}

 \newsavebox{\notrightarrow}
 \sbox{\notrightarrow}{$\to$\hspace{-4mm}/}
 
 \newsavebox{\PARTIALSLASH}
 \sbox{\PARTIALSLASH}{$\partial$\hspace{-1.6mm}/}
 
 \newsavebox{\ASLASH}
 \sbox{\ASLASH}{$A$\hspace{-2.1mm}/}
 
 \newsavebox{\KSLASH}
 \sbox{\KSLASH}{$k$\hspace{-1.8mm}/}
 
 \newsavebox{\LSLASH}
 \sbox{\LSLASH}{$\ell$\hspace{-1.8mm}/}
 
 \newsavebox{\QSLASH}
 \sbox{\QSLASH}{$q$\hspace{-1.8mm}/}
 
 \newsavebox{\DSLASH}
 \sbox{\DSLASH}{$D$\hspace{-2.2mm}/}
 
 \newsavebox{\DbfSLASH}
 \sbox{\DbfSLASH}{${\mathbf D}$\hspace{-2.8mm}/}
 
 \newsavebox{\DELVECRIGHT}
 \sbox{\DELVECRIGHT}{$\stackrel{\rightarrow}{\partial}$}
 
 \newcommand{\blue}{\IfColor{\textCadetBlue}{}}
\newcommand{\black}{\IfColor{\textBlack}{}}
\newcommand{\red}{\IfColor{\textRed}{}}
\newcommand{\green}{\IfColor{\textOliveGreen}{}}
\newcommand{\lil}{\IfColor{\textRedViolet}{}}

\newcommand{\bs}{\boldsymbol}

\makeatother
\usepackage[T1]{fontenc}
\usepackage[latin9]{inputenc}
\usepackage{amsmath}
\usepackage{amssymb}
\usepackage{graphicx}
\usepackage{dcolumn}
\usepackage{verbatim}
\usepackage{orcidlink}

\begin{document}
\title{Spontaneous breaking of global U(1) symmetry in\\ an interacting Bose gas under rigid rotation}
\author{E. Siri $^{a,b}$\,~\orcidlink{0009-0008-1021-3348}~~}\email{e.siri@physics.sharif.ir}\author{N. Sadooghi $^{a}$\,~\orcidlink{0000-0001-5031-9675}~~}\email{Corresponding author: sadooghi@physics.sharif.ir}
\affiliation{a) Department of Physics, Sharif University of Technology,
P.O. Box 11155-9161, Tehran, Iran}
\affiliation{b) Research Center for High Energy Physics, Department of Physics, Sharif University of Technology, P.O. Box 11155-9161, Tehran, Iran}
\begin{abstract}
We investigate the impact of rigid rotation on the spontaneous breaking of U(1) symmetry in a Bose gas, which is described by a self-interacting complex scalar field Lagrangian. Rigid rotation is introduced through a specific metric that explicitly depends on the angular velocity $\Omega$.  We begin by determining the free propagator for this model at finite temperature $T$ and chemical potential $\mu$. Using this propagator, we calculate the thermodynamic potential in terms of an energy dispersion relation $\epsilon_{k}$. It is found  that in both the U(1) symmetric phase and the symmetry-broken phase, two energy branches emerge. In the symmetry-broken phase, they are identified with a massive phonon and a massless roton mode. Notably, rotation does not alter $\epsilon_{k}$ at low momentum. Setting $\mu=0$, we use the total thermodynamic potential, which includes classical, thermal, vacuum, and nonperturbative ring contributions, to explore how the condensate depends on $T$ and $\Omega$. We first focus on the classical and thermal parts of the thermodynamic potential and find that the critical temperature of the U(1) phase transition scales as $\Omega^{1/3}$. By identifying the (pseudo-)Goldstone and non-Goldstone modes of this model with $\pi$ and $\sigma$ mesons, we calculate the $T$ and $\Omega$ dependence of masses $m_{\pi}$ and $m_{\sigma}$. We demonstrate that the Goldstone theorem holds only when the one-loop (thermal) corrections to $m_{\sigma}$ and $m_{\pi}$ are taken into account. We further explore the $T$ and $\Omega$ dependence of the condensate, determine the $\sigma$ dissociation temperatures for fixed $\Omega$, and compare them with the critical temperature of the phase transition. Additionally, we emphasize the role played by the nonperturbative ring potential, especially in altering the order of the phase transition with and without rotation.
\end{abstract}
\maketitle
\section{Introduction}\label{sec1}
\setcounter{equation}{0}
One of the primary goals of modern Heavy Ion Collision (HIC) experiments is to study matter under extreme conditions and its transitions through various phases. In Quantum Chromodynamics (QCD), these phases range from the deconfined quark-gluon plasma to the confined hadron phase, which consists of mesons and baryons. Mesons, as composite particles made up of a quark and an antiquark, are often regarded as (pseudo-)Goldstone bosons arising from the spontaneous breaking of chiral symmetry.
Key questions related to the phase transition of matter created in HIC experiments focus in particular on the order of the phase transition and the location of the critical endpoint \cite{hatsuda-book,fukushima2011,busza2018,bzdak2020,aarts2023}. Answers to these questions provide valuable insights into astrophysical and cosmological models of the early universe \cite{boyanovsky2006,laine-book}. Both of these properties are affected by external conditions, such as external electromagnetic fields and rotation. Intense magnetic fields are believed to be generated in the early stages of noncentral HICs. Depending on the initial conditions, the strength of the magnetic fields is estimated to be approximately $B\sim 10^{18}-10^{20}$ Gau\ss~in the early stages after these collisions \cite{skokov2009,shen2025}. In recent years, several studies have explored the QCD phase diagram in the presence of magnetic fields. Novel effects, such as magnetic and inverse magnetic catalysis are associated with the effect of constant background magnetic fields on the nature of the chiral phase transition and the location of the critical point \cite{fayazbakhsh2011,fayazbakhsh2012,fukushima2019}.
Recently, several studies have investigated the effect of rotation on quark matter created in HIC experiments. This matter is believed to experience extremely high vorticity, with an angular velocity reaching up to $10^{22}$ Hz \cite{becattini2017,becattini2020}. Extensive research has focused on how rotation influences the thermodynamic properties of relativistic fermionic systems \cite{yamamoto2013,chernodub2017a,chernodub2017b,ambrus2019,sadooghi2021,sun2024}. One notable example is the chiral vortical effect, which is related to the transport properties of the quark matter produced after HICs and provides insights into the topological aspects of QCD \cite{kharzeev2016}. When examining the thermodynamic properties of rotating Fermi gases using field theoretical methods, it is advantageous to assume rigid rotation with a constant angular velocity \cite{mameda2016, chen2016}. The impact of rigid rotation on QCD phase transitions, including chiral and confinement/deconfinement, has been studied with and without boundary conditions, e.g., in \cite{yamamoto2013, chernodub2021}.  In \cite{chernodub2021}, it is shown that at finite temperature the phase diagram of a uniformly rotating system exhibits, in addition to a confining and a deconfining phase at low and high temperatures, a mixed inhomogeneous phase at intermediate temperatures.
\par
Several studies have also explored both relativistic bosons \cite{braguta2023a,braguta2023b,ambrus2023,siri2024a,siri2024c,siri2024b,mameda2025,voskresensky2024a,voskresensky2024b,bordag2025a,bordag2025b} and the linear sigma model with quarks \cite{singha2024,hernandez2024,ambrus2025,singha2025} under rigid rotation. In \cite{braguta2023b}, a spin-one gluon gas under rigid rotation is analyzed, revealing that at temperatures below a certain supervortical temperature, the moment of inertia of a rotating spin-one gluon plasma becomes negative. This phenomenon indicates a thermodynamic instability and is associated with the negative Barnett effect, where the total angular moment of the system opposes the direction of its angular velocity. For spin-zero bosons in the presence of imaginary rotation, ninionic statistics arise, modifying the standard Bose-Einstein distribution with a statistical angle. Under specific conditions, these bosons exhibit fermionic-like behavior and display fractal thermodynamics that depend on the angle of imaginary rotation \cite{ambrus2023}. A separate study in \cite{siri2024a} investigated the thermodynamics of spin-zero complex scalar fields under rigid rotation, revealing that thermodynamic instabilities emerge at high temperatures and large coupling constants. These instabilities include negative moment of inertia and heat capacity.
Finally, in \cite{siri2024b}, the Bose-Einstein (BE) condensation of a free Bose gas subjected to rigid rotation is investigated in both relativistic and nonrelativistic limits. It is demonstrated that rotation not only modifies the equation of state of the system but also impacts the transition temperature for BEC and the fraction of condensates. Specifically, it is shown that the critical temperature of a rotating Bose gas is lower than that of a nonrotating gas; however, as the angular velocity increases, the critical temperature of the rotating gas also rises. Additionally, an analysis of the heat capacity of a nonrelativistic rotating free Bose gas indicates that rotation alters the nature of the BEC phase transition from continuous to discontinuous. The present paper aims to extend these findings to an interacting Bose gas under rigid rotation.
\par
We begin with the Lagrangian density of a complex Klein-Gordon field $\varphi$ that includes a self-interaction term $\lambda(\varphi^{\star}\varphi)^{2}$ with a coupling constant $\lambda$. To introduce rigid rotation we use a metric including the angular velocity $\Omega$. In the first part of this paper, we introduce a chemical potential $\mu$ corresponding to the global U(1) symmetry of the Lagrangian. For later analysis, we expand the Lagrangian density around a classical configuration $|\langle\varphi\rangle|\equiv v$. Following standard methods \cite{kapusta-book, schmitt-book-a} and utilizing an appropriate Bessel-Fourier transformation \cite{siri2024b,siri2024c}, we derive the free propagator of this model. This propagator is subsequently employed to compute the thermodynamic potential as a function of $\mu, \Omega$, and the energy dispersion relation $\epsilon_{k}^{\pm}$. As it turns out, the spontaneous breaking of U(1) symmetry occurs for $m<\mu$. In this regime, we find two distinct energy branches; one corresponding to a massive phonon and the other to a massless roton. It is noteworthy that the rotation does not alter  $\epsilon_{k}$ at low momentum, and the results are similar to the nonrotating case \cite{schmitt-book-b}.
\par
In the second part of this paper, we explore the impact of rotation on the spontaneous breaking of U(1) symmetry, focusing specifically on the case of zero chemical potential. Our primary emphasis is on the $T$ and $\Omega$ dependence of the critical temperature of the corresponding phase transition, as well as two masses $m_{1}$ and $m_{2}$, which are identified with the masses of the $\sigma$ and $\pi$ mesons, respectively. We begin by considering the thermodynamic potential discussed in the first part of this paper. Apart from a classical part, it consists of a thermal and a vacuum contributions. By employing a novel method for summing over the quantum number $\ell$ related to rotation, we perform a high-temperature expansion. Combining the classical and the thermal parts, we derive an analytical expression for the critical temperature of U(1) phase transition $T_{c}$, which is found to be proportional to $\Omega^{1/3}$. Furthermore, we show that the minima of this potential are proportional to $(1-t^{3})$, where $t\equiv T/T_{c}$ is the reduced temperature. This contrasts with the behavior observed in a nonrotating Bose gas, where the minima are described by the factor $(1-t_{0}^{2})$ with $t_{0}\equiv T/T_{c}^{(0)}$.\footnote{Here, sub- and superscripts zero correspond to nonrotating Bose gas.} We also demonstrate that when substituting these minima into $m_1$ and $m_2$, they become imaginary in the symmetry-restored phase, analogous to the behavior in a nonrotating Bose gas. This issue is addressed by adding the thermal masses that arise from one-loop perturbative contributions to $m_{1}$ and $m_{2}$. By following this method, we confirm that the Goldstone theorem is satisfied in the symmetry-restored phase.
\par
We then compute the vacuum part of the potential by adding the appropriate counterterms and performing  dimensional regularization. Our findings extend the results from \cite{carrington1992}, where the vacuum contribution to the effective action for a $\lambda\varphi^{4}$ theory was computed. We add this potential to the classical and thermal parts of the potential, minimize the resulting expression, and examine how the minima depend on temperature $T$ for fixed angular velocity $\Omega$. We show that, similar to the
behavior observed in a noninteracting Bose gas \cite{siri2024b}, rotation reduces the critical temperature of the phase transition, which then increases as $\Omega$ rises. Additionally, by plugging these minima into the corresponding expressions to $m_{1}$ and $m_{2}$ (or equivalently $m_{\sigma}$ and $m_{\pi}$), we investigate the $T$ dependence of $\sigma$ and $\pi$ meson masses for fixed $\Omega$. As expected, in the symmetry-restored phase, we find $m_{\sigma}=m_{\pi}$. This equality indicates that at $T_{c}$ the minima of the corresponding potential vanish, suggesting a second-order phase transition, even in the presence of rigid rotation.
\par
Finally, we focus on the nonperturbative ring contribution to the potential described above. We present a full derivation of the ring potential in the presence of rotation. Based on the findings in \cite{carrington1992}, we expect that the addition of the ring potential will alter the order of the phase transition. Our results indicate that when rotation is absent ($\Omega=0$), a discontinuous phase transition occurs at a specific temperature. In contrast, when rotation is present ($\Omega \neq 0$), the phase transition remains continuous. Furthermore, we define a $\sigma$ dissociation temperature, denoted by $T_{\text{diss}}$, which is characterized by $m_{\sigma}(T_{\text{diss}})=2m_{\pi}(T_{\text{diss}})$ and show that $T_{\text{diss}}$ is less than the critical temperature.
\par
The organization of this paper is as follows: In Sec. \ref{sec2}, we introduce the rigid rotation in the Lagrangian density of a complex scalar field in the presence of a finite chemical potential. We derive the corresponding free propagator, determine the full thermodynamic potential of this model, and explore how rotation affects the spontaneous breaking of global U(1) symmetry. In Sec. \ref{sec3}, we focus on the special case of $\mu=0$ and systematically determine the full thermodynamic potential, which consists, apart from the classical part, of a thermal and a vacuum contribution. After examining the effect of rotation on the Goldstone theorem, we add the nonperturbative ring contribution to this potential, which is explicitly derived for the case of a rotating complex scalar field. In Sec. \ref{sec4}, the numerically solve the corresponding gap equation for the full potential with and without the ring potential. We investigate the $T$ dependence of the corresponding minima for fixed $\Omega$. Additionally, we determine the $T$ and $\Omega$ dependence of $m_{\sigma}$ and $m_{\pi}$, along with the $\sigma$ dissociation temperatures.
Section \ref{sec5} concludes the paper with a compact summary of our findings. In Appendix \ref{appA}, we present the high-temperature expansion in the presence of a rigid rotation. Notably, we apply a method introduced in \cite{siri2024b} to sum over $\ell$. Appendices \ref{appB} and \ref{appC} contain derivations of formulas \eqref{E27} and \eqref{E34}, while the derivation of \eqref{E44} is detailed in Appendix \ref{appD}.
\section{Interacting charged scalars under rigid rotation}\label{sec2}
\subsection{The free propagator}\label{sec2A}
\setcounter{equation}{0}
We start with the Lagrangian density of a charged scalar field $\varphi$
\begin{eqnarray}\label{N1}
\mathscr{L}=g^{\mu\nu}\partial_{\mu}\varphi^{\star}\partial_{\nu}\varphi
-m^{2}\varphi^{\star}\varphi-\lambda(\varphi^{\star}\varphi)^{2},
\end{eqnarray}
with the metric
\begin{eqnarray}\label{N2}
g_{\mu \nu }=\left( \begin{matrix}
1-r^{2}\Omega^2 & y\Omega  & -x\Omega  & 0  \\
y\Omega  & -1 &0 & 0  \\
-x\Omega  & 0& -1 & 0  \\
0 & 0 & 0 & -1  \\
\end{matrix} \right),
\end{eqnarray}
describing a rigid rotation. Here, $m$ is the rest mass and $0<\lambda<1$ is the coupling constant, describing the strength of the interaction. The spacetime coordinate is described by  $x^{\mu}=(t,x,y,z)$ and $r^{2}\equiv x^{2}+y^{2}$. Moreover, $\Omega$ is the constant angular velocity of a rigid rotation around the $z$-axis.
The above Lagrangian is invariant under global $\mathrm{U}(1)$ transformation
\begin{eqnarray}\label{N3}
\varphi(x)\to e^{-i\alpha}\varphi(x),\qquad
\varphi^{\star}(x)\to e^{+i\alpha}\varphi^{\star}(x),
\end{eqnarray}
with $\alpha$ a real constant phase.
Plugging the metric into \eqref{N1}, we obtain
\begin{eqnarray}\label{N4}
\mathscr{L}=|(\partial_{0}-i\mu-i\Omega L_{z})\varphi|^{2}-|\boldsymbol{\nabla}\varphi|^{2}-m^{2}|\varphi|^{2}-\lambda|\varphi|^{4},
\end{eqnarray}
where the chemical potential $\mu$ corresponding to the global $\mathrm{U}(1)$ symmetry \eqref{N3} is introduced. The $z$-component of the angular momentum, $L_{z}$, is defined by $L_{z}=i\left(y\partial_{x}-x\partial_{y}\right)$.
To investigate the spontaneous breaking of U(1) symmetry, we rewrite $\mathscr{L}$ in terms of real fields $\varphi_{1}$ and $\varphi_{2}$ appearing in $\varphi=\frac{1}{\sqrt{2}}\left(\varphi_{1}+i\varphi_{2}\right)$ and perform the shift $\varphi_{i}\to \Phi_{i}+\varphi_{i}$ with $\Phi=\left(
\begin{array}{c}
v\\
0
\end{array}
\right)$ and $v=\text{const}$.
We arrive at
\begin{eqnarray}\label{N5}
\mathscr{L}=\sum_{i=0}^{4}\mathscr{L}_{i},
\end{eqnarray}
with
\begin{widetext}
\noindent
\begin{eqnarray}\label{N6}
\mathscr{L}_{0}&=&\frac{1}{2}(\mu^{2}-m^{2})v^{2}-\frac{\lambda}{4}v^{4},\nonumber\\
\mathscr{L}_{1}&=&(\mu^{2}-m^{2})v\varphi_{1}-\mu v\partial_{0}\varphi_{2}-\lambda v^{3}\varphi_{1}+i\mu\Omega v L_{z}\varphi_{2},\nonumber\\
\mathscr{L}_{2}&=&\frac{1}{2}\bigg\{(\partial_{0}\varphi_{1})^{2}+(\partial_{0}\varphi_{2})^{2}-(\bs{\nabla}\varphi_{1})^{2}-(\bs{\nabla}\varphi_{2})^{2}+(\mu^{2}-m^{2})\left(\varphi_{1}^{2}+\varphi_{2}^{2}\right)+2\mu\left(\varphi_{2}\partial_{0}\varphi_{1}-\varphi_{1}\partial_{0}\varphi_{2}\right)-\lambda\left(3v^{2}\varphi_{1}^{2}+v^{2}\varphi_{2}^{2}\right)\nonumber\\
&&~~~-\Omega^{2}\big[(L_{z}\varphi_{1})^{2}+(L_{z}\varphi_{2})^{2}\big]-2i\Omega\big[\left(\partial_{0}\varphi_{1}+\mu\varphi_{2}\right)L_{z}\varphi_{1}\big]-2i\Omega\big[\left(\partial_{0}\varphi_{2}-\mu\varphi_{1}\right)L_{z}\varphi_{2}\big]
\bigg\}, \nonumber\\
\mathscr{L}_{3}&=&-\lambda v\varphi_{1}\left(\varphi_{1}^{2}+\varphi_{2}^{2}\right), \nonumber\\
\mathscr{L}_{4}&=&-\frac{\lambda}{4}\left(\varphi_{1}^{2}+\varphi_{2}^{2}\right)^{2}.
\end{eqnarray}
\end{widetext}
The classical part of the Lagrangian, $\mathscr{L}_{0}$, defines the classical (zero mode) potential
\begin{eqnarray}\label{N7}
\mathcal{V}_{\text{cl}}(v)\equiv -\mathscr{L}_{0}=\frac{1}{2}(m^{2}-\mu^{2})v^{2}+\frac{\lambda}{4}v^{4}.
\end{eqnarray}
The free propagator arises from the quadratic term $\mathscr{L}_{2}$ in the fluctuating fields $\varphi_{1}$ and $\varphi_{2}$. To derive the free propagator in the momentum space, we use the Fourier-Bessel transformation
\begin{eqnarray}\label{N8}
\varphi_{i}(x)=\sqrt{\frac{\beta}{V}}\sum_{n,\ell,\bs{k}}~e^{i(\omega_{n}\tau+\ell \phi + k_{z}z)} J_{\ell}(k_{\perp}r) \tilde{\varphi}_{i}(k),\nonumber\\
\end{eqnarray}
with $i=1,2$. The cylindrical symmetry is implemented by introducing the cylinder coordinate system described by $x^{\mu}=(t,x,y,z)=\left(t,r\cos\phi,r\sin\phi,z\right)$, with $r$ the radial coordinate, $\phi$ the azimuthal angle, and $z$ the height of the cylinder. The conjugate momenta, corresponding to these coordinates at finite temperature $T$, are given by the bosonic Matsubara frequency $\omega_{n}=2\pi nT$, discrete quantum number $\ell$, which is the eigenvalue of $L_{z}$, continuous momentum $k_{z}$, and $k_{\perp}\equiv |\bs{k}_{\perp}|\equiv \left(k_{x}^{2}+k_{y}^{2}\right)^{1/2}$ in cylindrical coordinates. The Bessel function $J_{\ell}\left(k_{\perp}r\right)$ captures the radial dependence in this transformation and $\tau\equiv it$. Plugging \eqref{N8} into $\mathscr{L}_{2}$ and performing an integration over cylindrical coordinates, according to
\begin{eqnarray}\label{N9}
\int_{X}\equiv\int_{0}^{\beta}d\tau\int_{0}^{\infty}rdr\int_{0}^{2\pi}d\phi\int_{-\infty}^{\infty}dz,
\end{eqnarray}
we arrive after some manipulations at
\begin{widetext}
\begin{eqnarray}\label{N10}
\int_{X}\mathscr{L}_{2}=-\frac{V}{2}\sum\limits_{n,\ell,\bs{k}}
\left(\tilde{\varphi}_{1}(-k)~~\tilde{\varphi}_{2}(-k)\right)\left(\beta^{2}\mathcal{D}^{-1}_{\ell}(k)\right)
\left(
\begin{array}{c}
\tilde{\varphi}_{1}(k)\\
\tilde{\varphi}_{2}(k)
\end{array}
\right),
\end{eqnarray}
with the free propagator
\begin{eqnarray}\label{N11}
\beta^{2}\mathcal{D}_{\ell}^{-1}(k)=
\left(
\begin{array}{ccc}
(\omega_{n}+i\ell\Omega)^{2}+\omega_{1}^{2}-\mu^{2}&\qquad\qquad& -2\mu(\omega_{n}+i\ell\Omega)\\
2\mu\left(\omega_{n}+i\ell\Omega\right)&\qquad\qquad&(\omega_{n}+i\ell\Omega)^{2}+\omega_{2}^{2}-\mu^{2}
\end{array}
\right).
\end{eqnarray}
\end{widetext}
Here, $\omega_{i}^{2}\equiv \bs{k}^{2}+m_{i}^{2}, i=1,2$, with $m_{1}^{2}(v)\equiv 3\lambda v^{2}+m^{2}$ and $m_{2}^{2}(v)\equiv \lambda v^{2}+m^{2}$, the corresponding masses to two fields $\varphi_{1}$ and $\varphi_{2}$. In cylinder coordinate system, we have $\bs{k}^{2}\equiv \bs{k}_{\perp}^{2}+k_{z}^{2}$. In Sec. \ref{sec3}, we break the global $\mathrm{U}(1)$ symmetry by choosing $m^{2}=-c^{2}$ with $c^{2}>0$ and show that after considering the quantum corrections, $\varphi_{2}$ become a massless Goldstone mode.
\par
A comparison with similar results for a nonrotating charged Bose gas at $T$ and $\mu$ shows that while $\ell\Omega$ is said to play a role analogous to that of the chemical potential $\mu$ \cite{chen2016}, the manner in which it is incorporated into the free propagator and the thermodynamic potential differs significantly (as discussed below).
\subsection{The thermodynamic potential}\label{sec2B}
\begin{figure*}[hbt]
\includegraphics[width=8cm, height=6cm]{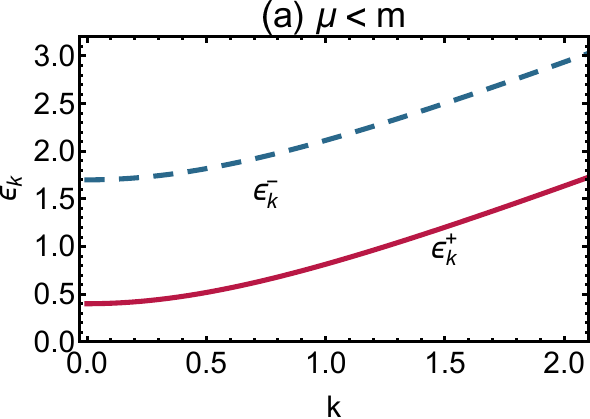}
\includegraphics[width=8cm, height=6cm]{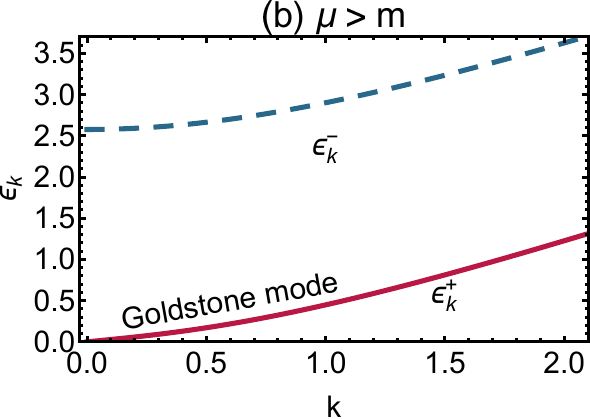}
\caption{(color online). The $k$ dependence of the energy dispersion $\epsilon_{k}^{\pm}$ from \eqref{N25} and \eqref{N26} in the $\mathrm{U}(1)$ symmetric phase (panel a) and the symmetry-broken phase (panel b), characterized by $\mu<m$ and $\mu>m$, respectively. As demonstrated, in the symmetry-broken phase, there is a massless Goldstone mode.  These findings remain unchanged regardless of any rotation.}\label{fig1}
\end{figure*}
To derive the thermodynamic potential $\mathcal{V}$, corresponding to this model, we follow the standard procedure and define this potential by
\begin{eqnarray}\label{N12}
\mathcal{V}=-\frac{T}{V}\ln \mathcal{Z},
\end{eqnarray}
with
\begin{eqnarray}\label{N13}
\ln \mathcal{Z}=-\frac{1}{2}\ln\det\left(\beta^{2}\mathcal{D}_{\ell}^{-1}(k)\right).
\end{eqnarray}
Let us first focus on $\ln\mathcal{Z}$ with $\mathcal{Z}$ the partition function of this model. Plugging $\mathcal{D}_{\ell}^{-1}$ from \eqref{N11} into \eqref{N13}, we arrive first at
\begin{eqnarray}\label{N14}
\hspace{-0.5cm}\ln \mathcal{Z}=-\frac{1}{2}\sum_{e=\pm}\sum_{n,\ell,\bs{k}}\ln\big|\beta^{2}[(\epsilon_{k}^{e})^{2}+(\omega_{n}+i
\ell\Omega)^{2}]\big|,
\end{eqnarray}
with $\epsilon^{\pm}_{k}$ given by
\begin{eqnarray}\label{N15}
\epsilon_{k}^{\pm}\equiv \left(E_{k}^{2}+\mu^{2}\mp\sqrt{4\mu^{2}E_{k}^{2}+\delta M^{4}}\right)^{1/2},
\end{eqnarray}
$E_{k}^{2}=\bs{k}^{2}+M^{2}$, and
\begin{eqnarray}\label{N16}
\hspace{-0.5cm}M^{2}\equiv\frac{1}{2}\left(m_{1}^{2}+m_{2}^{2}\right), \quad\delta M^{2}\equiv  \frac{1}{2}\left(m_{1}^{2}-m_{2}^{2}\right).
\end{eqnarray}
Following standard steps, it is possible to show that
\begin{eqnarray}\label{N17}
\ln \mathcal{Z}&=&-\frac{1}{4}\sum_{e=\pm}\sum_{n,\ell,\bs{k}}\bigg\{\ln\left(\beta^{2}[\omega_{n}^{2}+\left(\epsilon_{k}^{e}+\ell\Omega\right)^{2}]\right)\nonumber\\
&&+\ln\left(\beta^{2}[\omega_{n}^{2}+\left(\epsilon_{k}^{e}-\ell\Omega\right)^{2}]\right)\bigg\}.
\end{eqnarray}
Performing the Matsubara sum with
\begin{eqnarray}\label{N18}
\hspace{-0.5cm}\sum_{n=-\infty}^{+\infty}\ln\left((2\pi n)^{2}+\eta^{2}\right)=\eta+2\ln\left(1-e^{-\eta}\right),
\end{eqnarray}
we arrive at
\begin{eqnarray}\label{N19}
\ln \mathcal{Z}&=&-\frac{V}{2}\sum_{e=\pm}\sum_{\ell}\int d\tilde{k}~\bigg\{\beta \epsilon_{k}^{e}\nonumber\\
&&+\ln\left(1-e^{-\beta(\epsilon_{k}^{e}+\ell\Omega)}\right)+\ln\left(1-e^{-\beta(\epsilon_{k}^{e}-\ell\Omega)}\right)\bigg\},\nonumber\\
\end{eqnarray}
where the summation over $\bs{k}$ is replaced with the integration over $k$ in the cylinder coordinate system,
\begin{eqnarray}\label{N20}
\sum_{k}\to V\sum_{n,\ell}\int d\tilde{k},\quad\mbox{with}\quad
\int d\tilde{k}\equiv \int\frac{k_{\perp}dk_{\perp}dk_{z}}{(2\pi)^{3}}.\nonumber\\
\end{eqnarray}
Here, $k_{\perp}\equiv |\bs{k}_{\perp}|$. Using \eqref{N12}, the thermodynamic potential $\mathcal{V}$ is given by
\begin{eqnarray}\label{N21}
\mathcal{V}=\mathcal{V}_{\text{vac}}+\mathcal{V}_{T},
\end{eqnarray}
with the vacuum part
\begin{eqnarray}\label{N22}
\mathcal{V}_{\text{vac}}\equiv\frac{1}{2}\sum_{\ell}\int d\tilde{k}~\left(\epsilon_{k}^{+}+\epsilon_{k}^{-}\right),
\end{eqnarray}
and the matter (thermal) part
\begin{eqnarray}\label{N23}
\mathcal{V}_{T}&=&\frac{T}{2}\sum_{e=\pm}\sum_{\ell}\int d\tilde{k}~\bigg\{
\ln\left(1-e^{-\beta(\epsilon_{k}^{e}+\ell\Omega)}\right)\nonumber\\
&&
+\ln\left(1-e^{-\beta(\epsilon_{k}^{e}-\ell\Omega)}\right)\bigg\}.
\end{eqnarray}
Adding $\mathcal{V}$ with $\mathcal{V}_{\text{cl}}(v)$ from \eqref{N7}, to include the zero mode contribution, we obtain the full thermodynamic potential $\mathcal{V}_{\text{tot}}$,
\begin{eqnarray}\label{N24}
\mathcal{V}_{\text{tot}}&=&\frac{1}{2}(m^{2}-\mu^{2})v^{2}+\frac{\lambda}{4}v^{4}+
\frac{1}{2}\sum_{\ell}\int d\tilde{k}\left(\epsilon_{k}^{+}+\epsilon_{k}^{-}\right)
\nonumber\\
&&+\frac{T}{2}\sum_{e=\pm}\sum_{\ell\neq 0}\int d\tilde{k}~\bigg\{
\ln\left(1-e^{-\beta(\epsilon_{k}^{e}+\ell\Omega)}\right)\nonumber\\
&&
+\ln\left(1-e^{-\beta(\epsilon_{k}^{e}-\ell\Omega)}\right)\bigg\}.
\end{eqnarray}
\subsection{Spontaneous breaking of global U(1) symmetry}\label{sec2C}
Let us consider the classical potential \eqref{N7}. Assuming $m^{2}>\mu^{2}$, the coefficient of $v^{2}$ in this expression is positive and, as it turns out, $\mathcal{V}_{\text{cl}}$ possesses one single minimum at $\bar{v}_{0}=0$ and the system is in its symmetric phase. In this case, $m_{1}^{2}(\bar{v}_{0})=m_{2}^{2}(\bar{v}_{0})=m^{2}$, $\delta M^{2}=0$ and $\epsilon_{k}^{\pm}$ is given by
\begin{eqnarray}\label{N25}
\epsilon_{k}^{\pm}=\sqrt{\bs{k}^{2}+m^{2}}\mp \mu.
\end{eqnarray}
Here, $m$ is a mass gap and $\Delta\epsilon_{k}\equiv \epsilon_{k}^{-}-\epsilon_{k}^{+}=2\mu$. In Fig. \ref{fig1}(a), $\epsilon_{k}^{\pm}$ is plotted for generic mass $m=1$ MeV and chemical potential $\mu=0.6$ MeV ($\mu<m$).
\par
In the symmetry-broken phase characterized by $m^{2}<\mu^{2}$, however, extremizing $\mathcal{V}_{\text{cl}}$ yields a maximum at $v_{a}=0$ and two minima at
$$\bar{v}_{b}=\pm\sqrt{\frac{\mu^{2}-m^{2}}{\lambda}}.$$
The masses $m_{1}^{2}(\bar{v}_{b})=3\mu^{2}-2m^{2}$ and $m_{2}^{2}(\bar{v}_{b})=\mu^{2}$. We thus have $M^{2}=2\mu^{2}-m^{2}$ and $\delta M^{2}=\mu^{2}-m^{2}$ leading to
\begin{eqnarray}\label{N26}
\epsilon_{k}^{\pm}=\sqrt{\bs{k}^{2}+(3\mu^{2}-m^{2})\mp\sqrt{4\mu^{2}\bs{k}^{2}+(3\mu^{2}-m^{2})}}.\nonumber\\
\end{eqnarray}
In Fig. \ref{fig1}(b), $\epsilon_{k}^{\pm}$ is plotted for generic $\mu=1.1$ MeV and $m=1$ MeV ($\mu> m$). As it is shown, whereas $\epsilon_{k}^{-}$ is quadratic in $k\equiv|\bs{k}|$, $\epsilon_{k}^{+}\sim 0$ for $k\sim 0$. This behavior indicates the presence of a massless Goldstone mode. By expanding $\epsilon_{k}^{\pm}$ in the orders of $k\sim 0$, we obtain
\begin{eqnarray}\label{N27}
\epsilon_{k}^{-}&\simeq&\sqrt{2\left(3\mu^{2}-m^{2}\right)}+\frac{5\mu^{2}-m^{2}}{2\sqrt{2\left(3\mu^{2}-m^{2}\right)^{3}}}~\bs{k}^{2},\nonumber\\
\epsilon_{k}^{+}&\simeq&\sqrt{\frac{\mu^{2}-m^{2}}{3\mu^{2}-m^{2}}}~|\bs{k}|.
\end{eqnarray}
According to these results, $\epsilon_{k}^{+}$ and $\epsilon_{k}^{-}$ correspond to phonon and roton modes in the symmetry-broken phase $m<\mu$, respectively.
\par
 As it is shown in this section, $\ell\Omega$ appears in the thermal part of the effective potential $\mathcal{V}_{T}$ from \eqref{N23} and does not modify neither $m_{i}^{2}(v)$ nor the energy dispersion $\epsilon_{k}^{\pm}$.
Hence, a comparison with analogous results for nonrotating bosons \cite{schmitt-book-b} shows that rigid rotation has no effect on the behavior of $\epsilon_{k}^{\pm}$ at $k\sim 0$.
\subsection{Two special cases}\label{sec2D}
 In what follows, we consider two special cases $\lambda=0, \mu\neq 0$ and $\lambda\neq 0, \mu=0$:
\par
\textit{Case 1:} For the special case of noninteracting rotating Bose gas with $\lambda=0$ and $\mu\neq 0$, we have $m_{1}=m_{2}=m$, $E_{k}^{2}=\bs{k}^{2}+m^{2}$, and $\delta M=0$. We thus have
\begin{eqnarray}\label{N28}
\epsilon_{k}^{\pm}\big|_{\lambda=0,\mu\neq 0}=\sqrt{\bs{k}^{2}+m^{2}}\mp\mu,
\end{eqnarray}
and therefore
\begin{eqnarray}\label{N29}
\lefteqn{\mathcal{V}_{\text{tot}}\big|_{\lambda=0, \mu\neq 0}
=\frac{1}{2}(m^{2}-\mu^{2})v^{2}+
\sum_{\ell}\int d\tilde{k}~\bigg\{E_{k}
}\nonumber\\
&&+T\bigg[\ln\left(1-e^{-\beta(E_{k}-\mu_{\text{eff}})}\right)+\ln\left(1-e^{-\beta(E_{k}+\mu_{\text{eff}})}\right)\bigg]\bigg\},\nonumber\\
\end{eqnarray}
with $\mu_{\text{eff}}\equiv \mu+\ell\Omega$. This potential is exactly the same potential arising in \cite{siri2024b}. Using this potential, the effect of rotation on the BE condensation of a relativistic free Bose gas is studied.
\par
\textit{Case 2:} Another important case is characterized by $\lambda\neq 0$ and $\mu=0$. In this case, $\epsilon_{k}^{\pm}$ are given by
\begin{eqnarray}\label{N30}
\epsilon_{k}^{+}&=&\sqrt{\bs{k}^{2}+m_{2}^{2}}= \omega_{2}, \nonumber\\
\epsilon_{k}^{-}&=&\sqrt{\bs{k}^{2}+m_{1}^{2}}= \omega_{1}.
\end{eqnarray}
Plugging \eqref{N30} into \eqref{N24} and choosing $\mu=0$ and $m^{2}=-c^{2}$ with $c^{2}>0$, the total thermodynamic potential is given by
\begin{eqnarray}\label{N31}
\mathcal{V}_{\text{tot}}|_{\lambda\neq 0, \mu=0}=\mathcal{V}_{\text{cl}}+\mathcal{V}_{\text{vac}}+\mathcal{V}_{T},
\end{eqnarray}
with the classical part
\begin{eqnarray}\label{N32}
\mathcal{V}_{\text{cl}}=-\frac{c^{2}v^{2}}{2}+\frac{\lambda v^{4}}{4},
\end{eqnarray}
the vacuum part
\begin{eqnarray}\label{N33}
\mathcal{V}_{\text{vac}}=\frac{1}{2}\sum_{\ell}\int d\tilde{k}\left(\omega_{1}+\omega_{2}\right),
\end{eqnarray}
 and the thermal part
\begin{eqnarray}\label{N34}
\mathcal{V}_{T}=\frac{1}{2}\sum_{i=1,2}\left(\mathcal{V}_{i}^{+}+\mathcal{V}_{i}^{-}\right),
\end{eqnarray}
where
\begin{eqnarray}\label{N35}
\mathcal{V}_{i}^{\pm}\equiv T\sum_{\ell\neq 0}\int d\tilde{k}\ln\left(1-e^{-\beta(\omega_{i}\mp \ell\Omega)}\right).
\end{eqnarray}
Here, $\omega_{i},i=1,2$ are given in \eqref{N30}. Let us notice that in \eqref{N35}, the $\ell=0$ contribution is excluded, because the zero mode contribution is already captured by $\mathcal{V}_{\text{cl}}$ from \eqref{N32}. It is possible to limit the integration over $\ell$ in $\mathcal{V}_{T}^{\pm}$ from \eqref{N35}. Having in mind that the arguments of $\ln(1-e^{-\beta(\omega_{i}\mp \ell\Omega)})$ are to be positive, the summation over $\ell$ in $\ln(1-e^{-\beta(\omega_{i}- \ell\Omega)})$ is over $\ell\in (-\infty,-1]$ and in $\ln(1-e^{-\beta(\omega_{i}+ \ell\Omega)})$ is over $\ell\in [1,\infty)$ \cite{siri2024b}. Performing a change $\ell\to -\ell$, we thus have
\begin{eqnarray}\label{N36}
\hspace{-1cm}\mathcal{V}_{i}^{+}=T\sum_{\ell=1}^{\infty}\int d\tilde{k}~\ln\left(1-e^{-\beta(\omega_{i}+\ell\Omega)}\right)=\mathcal{V}_{i}^{-}.
\end{eqnarray}
Hence, the final form of $\mathcal{V}_{T}$ from \eqref{N34} reads
\begin{eqnarray}\label{N37}
\mathcal{V}_{T}=T\sum_{i=1,2}\sum_{\ell=1}^{\infty}\int d\tilde{k}\ln\left(1-e^{-\beta(\omega_{i}+\ell\Omega)}\right).
\end{eqnarray}
\section{Spontaneous breaking of global U(1) symmetry in a rigidly rotating Bose gas}\label{sec3}
\setcounter{equation}{0}
\subsection{The critical temperature of U(1) phase transition; analytical result}\label{sec3A}
In this section, we study the effect of rigid rotation on the spontaneous breaking of global $\mathrm{U}(1)$ symmetry in an interacting charged Bose gas. Before starting, we add a new term
\begin{eqnarray}\label{E1}
\widetilde{\mathscr{L}}_{0}=\frac{1}{2}m_{0}^{2}\left(\varphi_{1}+v\right)v,
\end{eqnarray}
to $\mathscr{L}$ from \eqref{N5}. This leads to an additional mass term in the classical potential $\mathcal{V}_{\text{cl}}$. We define a new mass $a^{2}\equiv c^{2}+m_{0}^{2}$, which replaces $c^{2}$ in \eqref{N32}. Minimizing the resulting expression, the (classical) minimum of $\mathcal{V}_{\text{cl}}$ is thus given by
\begin{eqnarray}\label{E2}
v_{0}^{2}\equiv \frac{a^{2}}{\lambda}.
\end{eqnarray}
At this minimum, the masses of $m_{1}^{2}(v)=3\lambda v^{2}-c^{2}$ and $m_{2}^{2}(v)=\lambda v^{2}-c^{2}$ are given by
\begin{eqnarray}\label{E3}
m_{1}^{2}(v_{0})=3a^{2}-c^{2}, \qquad m_{2}^{2}(v_{0})=a^{2}-c^{2}.
\end{eqnarray}
For $m_{0}=0$, we have $m_{2}=0$ and $\varphi_{2}$ becomes a massless Goldstone mode. The position of this (classical) minimum changes, once the contribution of the thermal part of the thermodynamic potential, $\mathcal{V}_{T}$, is considered. To show this, we first define $\mathcal{V}_{a}\equiv \mathcal{V}_{\text{cl}}+\mathcal{V}_{T}$ and use the high-temperature expansion of $\mathcal{V}_{T}$ by making use of the results presented in Appendix \ref{appA}. Considering only the first two terms of \eqref{appA13} and plugging the definitions of $m_{1}^{2}(v)$ and $m_{2}^{2}(v)$ into it, the high-temperature expansion of  $\mathcal{V}_{a}$ reads
\begin{eqnarray}\label{E4}
\mathcal{V}_{a}( v,T,\Omega)&=&-\frac{a^{2}v^{2}}{2}\left(1-\frac{2\lambda T^{3}\zeta(3)}{a^{2}\pi^{2}\Omega}\right)+\frac{\lambda v^{4}}{4}\nonumber\\
&&-\frac{2T^{5}\zeta(5)}{\pi^{2}\Omega}-\frac{c^{2}T^{3}\zeta(3)}{2\pi^{2}\Omega}+\cdots.
\end{eqnarray}
Setting the coefficient of $v^{2}$ equal to zero, the critical temperature of global $\mathrm{U}(1)$ phase transition is determined,
\begin{eqnarray}\label{E5}
T_{c}=\left(\frac{a^{2}\pi^{2}\Omega}{2\lambda\zeta(3)}\right)^{1/3}.
\end{eqnarray}
 In \cite{siri2024b}, the BE transition in a noninteracting Bose gas under rigid rotation is studied. It is shown that in nonrelativistic regime $T_{c}\propto \Omega^{2/5}$ and in ultrarelativistic regime $T_{c}\propto \Omega^{1/4}$. In the present case of interacting Bose gas, similar to that noninteracting cases, the critical temperature increases with increasing $\Omega$.
\par
\begin{figure}[hbt]
\includegraphics[width=8cm, height=6cm]{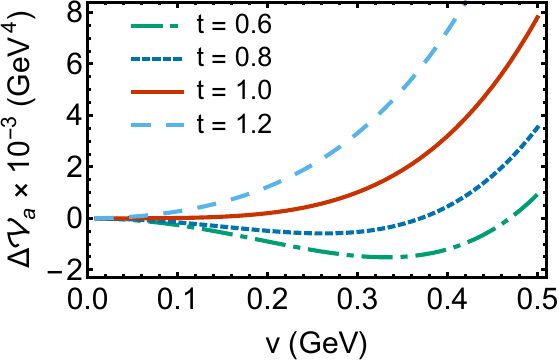}
\caption{(color online). The $v$ dependence of $\Delta\mathcal{V}_{a}$ from \eqref{E8} is plotted at $t=0.6,0.8,1,1.2$. At $t<1$ the global $\mathrm{U}(1)$ symmetry is broken and $\Delta\mathcal{V}_{a}$ possesses nontrivial minima at $v_{\min}^{2}=a^{2}(1-t^{3})/\lambda$. At $t=1$ the symmetry is restored and at $t\geq 1$ a single minimum at $v_{\text{min}}=0$ appears (see \eqref{E6}). }\label{fig2}
\end{figure}

Introducing the reduced temperature $t= T/T_{c}$, with $T_{c}=T_{c}(\Omega)$ from \eqref{E5}, and minimizing $\mathcal{V}_{a}$ from \eqref{E4} with respect to $v$, the new nontrivial minimum is given by
\begin{eqnarray}\label{E6}
v_{\text{min}}^{2}(T,\Omega)=
\begin{cases}
\dfrac{a^{2}}{\lambda}\left(1-t^{3}\right),&t< 1,\\
0,&t\geq 1.
\end{cases}
\end{eqnarray}
When comparing with a similar result for a nonrotating charged Bose gas \cite{kapusta-book}, it turns out that the power of $t$ in \eqref{E6} changes once the gas is subjected to small rotation.
In Sec. \ref{sec4}, we numerically study the effect of rotation on the spontaneous breaking of global $\mathrm{U}(1)$ symmetry. For this purpose, we employ a phenomenological model that includes $\sigma$ and $\pi$ mesons, replacing $\varphi_{1}$ and $\varphi_{2}$ fields in the above computation. We set $m_{1}^{2}(v_{0})=3\lambda v_{0}^{2}-c^{2}=m_{\sigma}^{2}$ and $m_{2}^{2}(v_{0})=\lambda v_{0}^{2}-c^{2}=m_{\pi}^{2}$ with $v_{0}$ the classical minimum from \eqref{E2}. Moreover, we choose $m_{0}$ in \eqref{E1} equal to $m_{\pi}$. For $m_{\sigma}=400$ MeV, and $m_{\pi}=140$ MeV, we obtain
\begin{eqnarray}\label{E7}
c=\left(\frac{m_{\sigma}^{2}-3m_{\pi}^{2}}{2}\right)^{1/2}\simeq 225~\text{MeV}.
\end{eqnarray}
Moreover, $a=\left(c^{2}+m_{\pi}^{2}\right)^{1/2}\simeq 265$ MeV. We also choose $\lambda=0.5$. Using these quantities the function
\begin{eqnarray}\label{E8}
\Delta\mathcal{V}_{a}&\equiv& \mathcal{V}_{a}(v,T,\Omega)-\mathcal{V}_{a}(0,T,\Omega)\nonumber\\
&=&-\frac{a^{2}v^{2}}{2}\left(1-t^{3}\right)+\frac{\lambda v^{4}}{4},
\end{eqnarray}
is plotted in Fig. \ref{fig2} at $t=0.6,0.8$ in the symmetry-broken phase and $t=1.2$ in the symmetry-restored phase. At $t=1$ a phase transition from the symmetry-broken phase to a symmetry-restored phase occurs. Let us notice, that the effect of rotation consists of changing the power of $t$ in \eqref{E6} and \eqref{E8} from $t^{2}$ to $t^{3}$.  This is apart from the $\Omega$ dependence of the critical temperature $T_{c}$ from \eqref{E5} (see Fig. \ref{fig7}).
\par
The result indicates a continuous phase transition from a symmetry-broken phase at $t<1$ to a symmetry-restored phase at $t\geq 1$. To scrutinize this conclusion, let us consider the pressure $P$ arising from $\mathcal{V}_{a}$ from \eqref{E4}. It is given by $P=-\mathcal{V}_{a}$. Denoting the pressures below and above $T_{c}$ with $P_{<}(v,T,\Omega)$ and $P_{>}(v,T,\Omega)$, we
have
\begin{eqnarray}\label{E9}
P_{<}\left(v_{\text{min}},T,\Omega\right)&=&-\frac{a^{4}}{2\lambda}t^{3}+\frac{c^{2}T^{3}\zeta(3)}{2\pi^{2}\Omega}+\frac{2T^{5}\zeta(5)}{\pi^{2}\Omega}+\frac{a^{4}}{4\lambda}t^{6},\nonumber\\
P_{>}\left(0,T,\Omega\right)&=&\frac{c^{2}T^{3}\zeta(3)}{2\pi^{2}\Omega}+\frac{2T^{5}\zeta(5)}{\pi^{2}\Omega}-\frac{a^{4}}{4\lambda}.
\end{eqnarray}
Here, we have added a term $-a^{4}/4\lambda$ to $P_{<}$ and $P_{>}$ in order to guarantee $P_{<}(v_{\text{min}}^{2},0,\Omega)=0$ and $P_{<}=P_{>}$ at the the transition temperature $T_{c}$. At $T=T_{c}$, the pressure is given by
\begin{eqnarray}\label{E10}
P_{<}\left(v_{\text{min}}^{2},T_{c},\Omega\right)&=&P_{>}\left(0,T_{c},\Omega\right)\nonumber\\
&=&-\frac{a^{4}}{4\lambda}+\frac{a^{2}c^{2}}{4\lambda}+\frac{a^{10/3}\pi^{4/3}\Omega^{2/3}\zeta(5)}{2^{2/3}\lambda^{5/3}[\zeta(3)]^{5/3}}. \nonumber\\
\end{eqnarray}
For $m_{0}=0$ (or $a=c$), the first two terms cancel, resulting in an increase in pressure as $\Omega$ increases. Moreover, whereas the entropy ($dP/dT$) is continuous at $T=T_{c}$,
\begin{eqnarray}\label{E11}
\frac{dP_{<}}{dT}\bigg|_{T_{c}}=\frac{dP_{>}}{dT}\bigg|_{T_{c}},
\end{eqnarray}
the heat capacity ($d^{2}P/dT^{2}$) is discontinuous
\begin{eqnarray}\label{E12}
\frac{d^{2}P_{<}}{dT^{2}}\bigg|_{T_{c}}-\frac{d^{2}P_{>}}{dT^{2}}\bigg|_{T_{c}}=\frac{ 9 c^{8/3} [\zeta (3)]^{2/3}}{2^{1/3} \pi ^{4/3} \lambda^{1/3} \Omega ^{2/3}}.
\end{eqnarray}
Hence, according to Ehrenfest classification, this is a second order phase transition. In comparison to the nonrotating case \cite{kapusta-book}, although rotation alters the critical temperature, the order of the phase transition remains unchanged. It is noteworthy that the discontinuity in the heat capacity decreases with increasing $\Omega$.
\par
Plugging at this stage, $v_{\text{min}}^{2}$ from \eqref{E6} into $m_{1}^{2}(v)=3\lambda v^{2}-c^{2}$ and $m_{2}^{2}(v)=\lambda v^{2}-c^{2}$, we arrive at
\begin{eqnarray}\label{E13}
m_{1}^{2}(v_{\text{min}})&=&
\begin{cases}
3a^{2}(1-t^3)-c^{2},&t<1,\\
-c^{2},&t\geq 1,
\end{cases}\nonumber\\
m_{2}^{2}(v_{\text{min}})&=&
\begin{cases}
a^{2}(1-t^3)-c^{2},&~~~t<1,\\
-c^{2},&~~~t\geq 1.
\end{cases}\nonumber\\
\end{eqnarray}
Hence, as it turns out, at $t\geq 1$, after the symmetry is restored, $m_{1}^{2}$ and $m_{2}^{2}$ become negative. Contrary to our expectation, for $a=c$, i.e. in the chiral limit $m_{0}=0$, the Goldstone boson $\varphi_{2}$ acquires a negative mass $-c^{2}t^{3}$ in the symmetry-broken phase at $t<1$. In what follows, we compute the one-loop tadpole diagram contributions to masses $m_{1}$ and $m_{2}$. We show, in particular, that by considering the thermal mass, the one-loop corrected mass of the Goldstone mode $\varphi_{2}$ vanishes in chiral limit $m_{0}=0$.
\subsection{One-loop corrections to $\bs{m_{1}(v)}$ and $\bs{m_{2}(v)}$}\label{sec3B}
\begin{figure}[hbt]
\includegraphics[width=8cm, height=2.3cm]{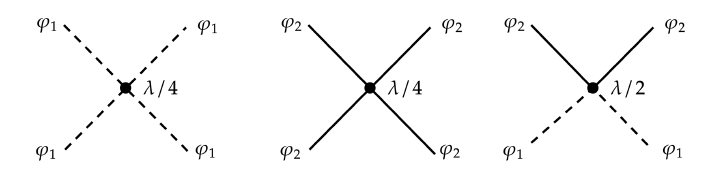}
\caption{Three vertices arising from $\mathscr{L}_{4}$ from \eqref{N4}. Dashed and solid lines correspond to $\varphi_{1}$ and $\varphi_{2}$ fields, respectively. }\label{fig3}
\end{figure}

To calculate the one-loop corrections to $m_{1}$ and $m_{2}$, let us consider $\mathscr{L}_{4}$ from \eqref{N4}. Three vertices, corresponding to three terms in $\mathscr{L}_{4}=-\frac{\lambda}{4}\left(\varphi_{1}^{2}+\varphi_{2}^{2}\right)^{2}$, are to be considered in this computation (see Fig. \ref{fig3}),
\begin{eqnarray}\label{E14}
-\frac{\lambda}{4}\varphi_{1}^{4}&\to& -\frac{\lambda}{4},\nonumber\\
-\frac{\lambda}{4}\varphi_{2}^{4}&\to& -\frac{\lambda}{4},\nonumber\\
-\frac{\lambda}{2}\varphi_{1}^{2}\varphi_{2}^{2}&\to&-\frac{\lambda}{2}.
\end{eqnarray}
\par
\begin{figure*}[hbt]
\includegraphics[width=12.5cm, height=2.8cm]{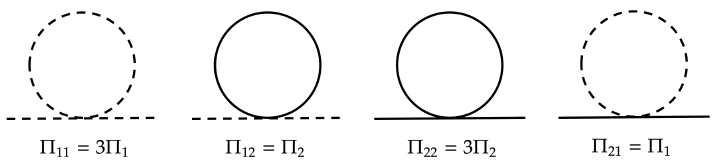}
\caption{ The tadpole diagrams contributing to the one-loop corrections of $m_{1}^{2}$ and $m_{2}^{2}$. Dashed and solid lines correspond to $\varphi_{1}$ and $\varphi_{2}$ fields, respectively. }\label{fig4}
\end{figure*}

They lead to two different tadpole contributions to $\langle\Omega| T\left(\varphi_{1}(x)\varphi_{1}(y)\right)|\Omega\rangle$ and  $\langle\Omega| T\left(\varphi_{2}(x)\varphi_{2}(y)\right)|\Omega\rangle$ that correct $m_{1}$ and $m_{2}$ perturbatively. They are denoted by $\Pi_{ij}$ with the first index, $i=1,2$, corresponds to whether $\varphi_{1}$ or $\varphi_{2}$ are in the external legs, and the second index $j=1,2$ to whether the internal loop is built from $\varphi_{1}$ or $\varphi_{2}$ (see Fig. \ref{fig4}, where $\Pi_{ij}$ are plotted). Hence, according to this notation, the one-loop perturbative corrections to $m_{1}^{2}$ and $m_{2}^{2}$ arise from
\begin{eqnarray}\label{E15}
m_{1}^{2}(v)&\to& m_{1}^{2}(v)+\Pi_{11}+\Pi_{12},\nonumber\\
m_{2}^{2}(v)&\to& m_{2}^{2}(v)+\Pi_{21}+\Pi_{22}.
\end{eqnarray}
At this stage, we introduce
\begin{eqnarray}\label{E16}
\Pi_{i}(T,\Omega,m_{i})\equiv \lambda T\sum_{n=-\infty}^{\infty}\sum_{\ell=-\infty}^{\infty}\int d\tilde{k}~D_{\ell}\left(\omega_{n},\omega_{i}\right),\nonumber\\
\end{eqnarray}
with free boson propagator
\begin{eqnarray}\label{E17}
D_{\ell}\left(\omega_{n},\omega_{i}\right)\equiv
\frac{1}{\left(\omega_{n}-i\ell\Omega\right)^{2}+\omega_{i}^{2}},
\end{eqnarray}
arising from \eqref{N11} with $\mu=0$. Here, $\omega_{i}^{2}=\bs{k}_{\perp}^{2}+k_{z}^{2}+m_{i}^{2}$ and $i=1,2$. Using this notation, it turns out that
\begin{eqnarray}\label{E18}
\Pi_{11}=3\Pi_{1}, &\qquad&\Pi_{12}=\Pi_{2},\nonumber\\
\Pi_{22}=3\Pi_{2},&\qquad&\Pi_{21}=\Pi_{1}.
\end{eqnarray}
Hence, the perturbative corrections of masses are given by
\begin{eqnarray}\label{E19}
m_{1}^{2}(v)&\to& m_{1}^{2}(v)+3\Pi_{1}+\Pi_{2},\nonumber\\
m_{2}^{2}(v)&\to& m_{2}^{2}(v)+3\Pi_{2}+\Pi_{1}.
\end{eqnarray}
To evaluate $\Pi_{i}$ from \eqref{E16}, we follow the same steps as presented in \cite{siri2024b}.
The Matsubara summation is evaluated with
\begin{eqnarray}\label{E20}
\sum_{n}D_{\ell}\left(\omega_{n},\omega_{i}\right)=\frac{1}{2T\omega_{i}}[n_{b}\left(\omega_{i}+\ell\Omega\right)+n_{b}\left(\omega_{i}-\ell\Omega\right)+1], \nonumber\\
\end{eqnarray}
where $n_{b}(\omega)\equiv 1/\left(e^{\beta\omega}-1\right)$ is the BE distribution function. In what follows, we insert \eqref{E20} into \eqref{E16} and focus only on the matter ($T$ and $\Omega$ dependent) part of $\Pi_{i}$,
\begin{eqnarray}\label{E21}
\Pi_{i}^{\text{mat}}=\frac{\lambda}{2}\sum_{e=\pm}\sum_{\ell\neq 0}\int d\tilde{k}~\frac{n_{b}(\omega_{i}+e\ell\Omega)}{\omega_{i}}.
\end{eqnarray}
Having in mind that in $n_{b}(\omega_{i}\pm \ell\Omega)$, we must have $e^{\beta(\omega_{i}\pm\ell\Omega)}-1>0$, it is possible to limit the summation over $\ell$. We thus obtain
\begin{eqnarray}\label{E22}
\Pi_{i}^{\text{mat}}=\lambda\sum_{\ell=1}^{\infty}\int d\tilde{k}~\frac{n_{b}(\omega_{i}+\ell\Omega)}{\omega_{i}}.
\end{eqnarray}
Let us notice that in the term including $n_{b}(\omega_{i}-\ell\Omega)$ an additional shift $\ell\to -\ell$ is performed.
To carry out the summation over $\ell$ and eventually the integration over $k_{\perp}$ and $k_{z}$, we use
\begin{eqnarray}\label{E23}
n_{b}(\omega_{i}+\ell\Omega)=T\frac{d}{d\omega_{i}}\ln\left(1-e^{-\beta(\omega_{i}+\ell\Omega)}\right),
\end{eqnarray}
and arrive first at
\begin{eqnarray}\label{E24}
\Pi_{i}^{\text{mat}}=\lambda T\sum_{\ell=1}^{\infty}\int d\tilde{k}~\frac{1}{\omega_{i}}\frac{d}{d\omega_{i}}\ln\left(1-e^{-\beta(\omega_{i}+\ell\Omega)}\right). \nonumber\\
\end{eqnarray}
Using, at this stage, \eqref{appA2}, we then obtain
\begin{eqnarray}\label{E25}
\Pi_{i}^{\text{mat}}&=&-\lambda T\sum_{\ell=1}^{\infty}\sum_{j=1}^{\infty}\int d\tilde{k}~\frac{1}{\omega_{i}}\frac{d}{d\omega_{i}}\frac{e^{-\beta\omega_{i}j}e^{-\beta\ell\Omega j}}{j}\nonumber\\
&=&\lambda\sum_{\ell=1}^{\infty}\sum_{j=1}^{\infty}\int d\tilde{k}~\frac{e^{-\beta\omega_{i}j}e^{-\beta\ell\Omega j}}{\omega_{i}}.
\end{eqnarray}
The summation over $\ell$ can be performed by making use of \eqref{appA4}.  Assuming $\beta\Omega<1$ and using \eqref{appA5}, $\Pi_{i}^{\text{mat}}$ reads
\begin{eqnarray}\label{E26}
\Pi_{i}^{\text{mat}}=\frac{\lambda}{\beta\Omega}\sum_{j=1}^{\infty}\frac{1}{j}\int d\tilde{k}~\frac{e^{-\beta \omega_{i}j}}{\omega_{i}}.
\end{eqnarray}
Following the method presented in Appendix \ref{appB}, we finally arrive at
\begin{eqnarray}\label{E27}
\Pi_{i}^{\text{mat}}=\frac{\lambda T^{3}\zeta(3)}{2\pi^{2}\Omega}+\cdots.
\end{eqnarray}
The first term in \eqref{E27} is analogous to the thermal mass $\lambda T^{2}/3$ in a nonrotating interacting Bose gas \cite{kapusta-book} and the ellipsis includes higher order corrections of $\Pi_{i}^{\text{mat}}$ in $\beta m_{i}$. At high temperature, it is enough to consider only the first term in \eqref{E27}, which is independent of $m_{i}$. We thus have
\begin{eqnarray}\label{E28}
\Pi_{1}^{\text{mat}}=\Pi_{2}^{\text{mat}}=\frac{\lambda T^{3}\zeta(3)}{2\pi^{2}\Omega},
\end{eqnarray}
and therefore
\begin{eqnarray}\label{E29}
m_{1}^{2}(v)&\to& m_{1}^{2}(v)+4\Pi_{1}^{\text{mat}}=m_{1}^{2}(v)+a^{2}t^{3},\nonumber\\
m_{2}^{2}(v)&\to& m_{2}^{2}(v)+4\Pi_{2}^{\text{mat}}=m_{2}^{2}(v)+a^{2}t^{3},\nonumber\\
\end{eqnarray}
with $t=T/T_{c}$ and $T_{c}$ from \eqref{E5}.
\subsection{Goldstone theorem}\label{sec3C}
\begin{figure}[hbt]
\includegraphics[width=8cm, height=6cm]{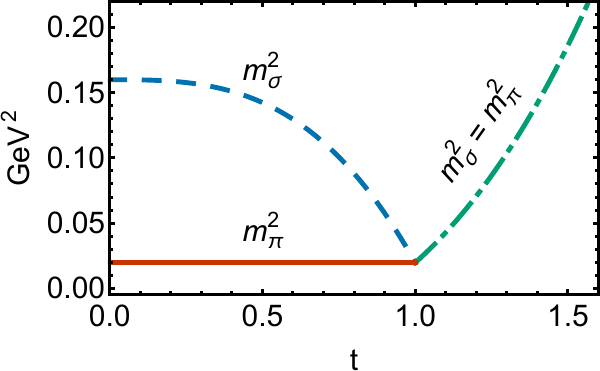}
\caption{(color online). The $t$ dependence of $m_{1}^{2}$ and $m_{2}^{2}$ from \eqref{E30} at $v_{\text{min}}^{2}$ from \eqref{E6} is plotted. These masses are identified with $\sigma$ and $\pi$ meson masses. In the symmetry-broken phase, at $t<1$, $m_{\sigma}$ decreases with increasing temperature, while $m_{\pi}$ remains constant. At symmetry-restored phase at $t\geq 1$, $m_{\sigma}$ and $m_{\pi}$ are equal and increase with increasing $t$.}\label{fig5}
\end{figure}

Let us consider again the result presented in \eqref{E13}. Adding the contribution of thermal mass \eqref{E28} to $m_{1}^{2}(v_{\text{min}}^{2})$ and $m_{2}^{2}(v_{\text{min}}^{2})$, according to \eqref{E29}, we obtain
\begin{eqnarray}\label{E30}
m_{1}^{2}(v_{\text{min}})&=&
\begin{cases}
2c^{2}(1-t^{3})+3m_{0}^{2}\left(1-\dfrac{2t^{3}}{3}\right),&t<1,\\
c^{2}(t^{3}-1)+m_{0}^{2}t^{3},&t\geq 1,
\end{cases}\nonumber\\
m_{2}^{2}(v_{\text{min}})&=&
\begin{cases}
m_{0}^{2},&~~~\qquad\qquad t<1,\\
c^{2}(t^{3}-1)+m_{0}^{2}t^{3},&~~~\qquad\qquad t\geq 1,
\end{cases}\nonumber\\
\end{eqnarray}
where $a^{2}=c^{2}+m_{0}^{2}$ is used.  Assuming $m_{0}=0$, $m_{2}$ vanishes at $t<1$. This indicates that the Goldstone theorem is valid when the thermal mass corrections to $m_{1}^{2}$ and $m_{2}^{2}$ are taken into account. Moreover, we observe that $m_{1}^{2}(v_{\text{min}})=m_{2}^{2}(v_{\text{min}})$ in the symmetry-restored phase at $t\geq 1$.  In Fig. \ref{fig5}, the $t$ dependence of  $m_{1}^{2}(v_{\text{min}})$ and $m_{2}^{2}(v_{\text{min}})$ from \eqref{E30} is plotted. These masses are identified with $m_{\sigma}^{2}$ and $m_{\pi}^{2}$, respectively. We use $c\simeq 0.225$ GeV from \eqref{E7} and $m_{0}=0.140$ GeV, as described in Sec. \ref{sec3B} and observe that in the symmetry-broken phase, at $t<1$, $m_{\sigma}$ decreases with increasing temperature, while $m_{\pi}$ remains constant. As expected, at symmetry-restored phase at $t\geq 1$, $m_{\sigma}$ and $m_{\pi}$ are equal and increase with increasing temperature. It is noteworthy that the effect of rotation, apart from affecting the value of the critical temperature $T_{c}$ from \eqref{E5}, consists of changing the power of $t$ in \eqref{E30} from $t^{2}$ to $t^{3}$ (see \cite{kapusta-book}).
\subsection{Vacuum potential}\label{sec3D}
In what follows, we compute the contribution of the vacuum part of the thermodynamic potential, $\mathcal{V}_{\text{vac}}$ from \eqref{N33} to $\mathcal{V}_{\text{tot}}$. Let us first consider the summation over $\ell\in (-\infty,+\infty)$ in this expression. This sum is divergent and need an appropriate regularization. To perform the summation over $\ell$, we use
\begin{eqnarray}\label{E31}
\sum_{\ell=-\infty}^{\infty}1&=&\lim\limits_{x\to 0}\sum_{\ell=-\infty}^{\infty}e^{-\ell^{2}x}\nonumber\\
&=&\lim\limits_{x\to 0}\left(1+2\sum_{\ell=1}^{\infty}e^{-\ell^{2}x}\right)=1+\lim\limits_{x\to 0}\frac{1}{1-e^{-x}}\nonumber\\
&=&1+\mbox{divergent term}.
\end{eqnarray}
Neglecting the divergent term, we obtain
\begin{eqnarray}\label{E32}
\mathcal{V}_{\text{vac}}=\frac{1}{2}\int d\tilde{k}\left(\omega_{1}+\omega_{2}\right).
\end{eqnarray}
The above regularization guarantees that rotation does not alter $\mathcal{V}_{\text{vac}}$. To perform the integration over $k_{\perp}$ and $k_{z}$, let us consider the integral
\begin{eqnarray}\label{E33}
I(m)\equiv \frac{\bar{\mu}^{\epsilon}}{2}\int d\tilde{k}\left(\bs{k}^{2}+m^{2}\right)^{1/2},
\end{eqnarray}
with $\epsilon=3-d$. Here, $d$ is the dimension of spacetime and $\bar{\mu}$ denotes an appropriate energy scale. Later, we show that $\bar{\mu}$ can be eliminated from the computation. Utilizing
\begin{eqnarray}\label{E34}
\hspace{-1cm}\Phi(m,d,n)&=&\int\frac{d^{d}k}{(2\pi)^{d}}\frac{1}{\left(\bs{k}^{2}+m^{2}\right)^{n}}\nonumber\\
&=&\frac{1}{(4\pi)^{d/2}}\frac{\Gamma\left(n-d/2\right)}{\Gamma(n)}\frac{1}{\left(m^{2}\right)^{n-d/2}},
\end{eqnarray}
to perform a $d$ dimensional regularization, we obtain for $\Phi(m,3-\epsilon,-1/2)$,\footnote{In Appendix \ref{appC}, we derive \eqref{E24} in cylinder coordinate system.}
\begin{eqnarray}\label{E35}
\hspace{-0.2cm}I(m)=-\frac{m^{4}}{64\pi^{2}}\left(\frac{2}{\epsilon}+\frac{3}{2}-\gamma_{E}-\ln\frac{m^{2}}{4\pi \bar{\mu}^{2}}\right).
\end{eqnarray}
The vacuum part of the thermodynamic potential \eqref{E32} is thus given by
\begin{eqnarray}\label{E36}
\mathcal{V}_{\text{vac}}&=&I(m_{1})+I(m_{2})\nonumber\\
&=&-\frac{(m_{1}^{4}+m_{2}^{4})}{64\pi^{2}}\left(\frac{2}{\epsilon}+\frac{3}{2}-\gamma_{E}\right)+\frac{m_{1}^{4}}{64\pi^{2}}\ln\frac{m_{1}^{2}}{4\pi\bar{\mu}^{2}}
\nonumber\\
&&+\frac{m_{2}^{4}}{64\pi^{2}}\ln\frac{m_{2}^{2}}{4\pi\bar{\mu}^{2}}.
\end{eqnarray}
In what follows, we regularize this potential by following the method presented in \cite{carrington1992}. To do this, we first define
\begin{eqnarray}\label{E37}
\mathcal{V}_{b}\equiv \mathcal{V}_{\text{cl}}+\mathcal{V}_{\text{vac}}+\mathcal{V}_{\text{CT}},
\end{eqnarray}
with $\mathcal{V}_{\text{cl}}$ from \eqref{N32} with $c^{2}$ replaced with $a^{2}=c^{2}+m_{0}^{2}$ and $\mathcal{V}_{\text{vac}}$ from \eqref{E36}. The counterterm potential is given by
\begin{eqnarray}\label{E38}
\mathcal{V}_{\text{CT}}=\frac{Av^{2}}{2}+\frac{B v^{4}}{4}+C.
\end{eqnarray}
The coefficients $A$ and $B$ are determined by utilizing two prescriptions
\begin{eqnarray}\label{E39}
\frac{\partial\mathcal{V}_{b}}{\partial v}\bigg|_{v_{0}^{2}}=0,\qquad
\frac{\partial^{2}\mathcal{V}_{b}}{\partial v^{2}}\bigg|_{v_{0}^{2}}=m_{1}^{2}(v_{0}).
\end{eqnarray}
 Here, $v_{0}^{2}$ from \eqref{E2} is the classical minimum and $m_{1}^{2}(v_{0})$ from \eqref{E3}. Let us note that the first prescription guarantees that the position of the classical minimum does not change by considering the vacuum part of the potential. The term $C$ in \eqref{E38} includes all terms which are independent of $v$. Using \eqref{E39}, we arrive at
\begin{eqnarray}\label{E40}
A&=&-\frac{m_{0}^{2}}{2}+\frac{3c^{2}\lambda}{8\pi^{2}}
+\frac{c^{2}\lambda\gamma_{E}}{4\pi^{2}}+\frac{5m_{0}^{2}\lambda}{8\pi^{2}}
-\frac{c^{2}\lambda}{2\pi^{2}\epsilon}\nonumber\\
&&+\frac{c^{2}\lambda}{16\pi^{2}}\ln\left(\frac{m_{0}^{2}}{4\pi\bar{\mu}^{2}}\right)+\frac{3c^{2}\lambda}{16\pi^{2}}\ln\left(\frac{2c^{2}+3m_{0}^{2}}{4\pi\bar{\mu}^{2}}\right),\nonumber\\
B&=&\frac{m_{0}^{2}\lambda}{2a^{2}}-\frac{5\lambda^{2}\gamma_{E}}{8\pi^{2}}+\frac{5\lambda^{2}}{4\pi^{2}\epsilon}-\frac{\lambda^{2}}{16\pi^{2}}\ln\left(\frac{m_{0}^{2}}{4\pi\bar{\mu}^{2}}\right)\nonumber\\
&&-\frac{9\lambda^{2}}{16\pi^{2}}\ln\left(\frac{2c^{2}+3m_{0}^{2}}{4\pi\bar{\mu}^{2}}\right).
\end{eqnarray}
Plugging $A$ and $B$ from \eqref{E40} into $\mathcal{V}_{\text{CT}}$ from \eqref{E38} and choosing
\begin{eqnarray}\label{E41}
C&=&\frac{c^{4}}{16\pi^{2}\epsilon}-\frac{c^{4}}{64\pi^{2}}\ln\left(\frac{m_{0}^{2}}{4\pi\bar{\mu}^{2}}\right)-\frac{c^{4}}{64\pi^{2}}\ln\left(\frac{2c^{2}+3m_{0}^{2}}{4\pi\bar{\mu}^{2}}\right)\nonumber\\
&&+\frac{3c^{4}}{64\pi^{2}}-\frac{c^{4}\gamma_{E}}{32\pi^{2}},
\end{eqnarray}
the counterterm potential from \eqref{E38} is determined. These counterterms eliminate the divergent terms in the vacuum potential, as expected. The total potential $\mathcal{V}_{b}$ from \eqref{E37} is thus given by
\begin{eqnarray}\label{E42}
\mathcal{V}_{b}&=&-\frac{a^{2}v^{2}}{2}+\frac{\lambda v^{4}}{4}
-\frac{m_{0}^{2}v^{2}}{4}+\frac{3c^{2}\lambda v^{2}}{8\pi^{2}}+\frac{5m_{0}^{2}\lambda v^{2}}{16\pi^{2}}
\nonumber\\
&&-\frac{15\lambda^{2}v^{4}}{64\pi^{2}}+\frac{m_{0}^{2}\lambda v^{4}}{8a^{2}}\nonumber\\
&&+
\frac{m_{1}^{4}}{64\pi^{2}}\ln\left(\frac{m_{1}^{2}}{2c^{2}+3m_{0}^{2}}\right)+\frac{m_{2}^{4}}{64\pi^{2}}\ln\left(\frac{m_{2}^{2}}{m_{0}^{2}}\right). \nonumber\\
\end{eqnarray}
As mentioned earlier, the energy scale $\bar{\mu}$ does not appear in the final expression of $\mathcal{V}_{b}$. Additionally, a nonzero $m_{0}$ is necessary to specifically regularize the last term in $\mathcal{V}_{b}$ from \eqref{E42}.
\subsection{Ring potential}\label{sec3E}
\begin{figure*}[hbt]
\includegraphics[width=0.9\textwidth]{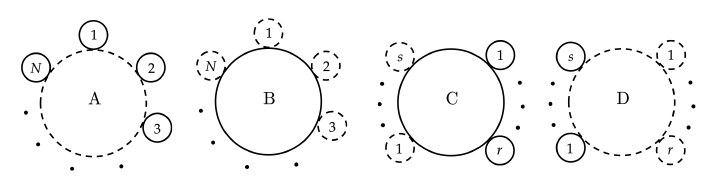}
\caption{Ring diagrams of Type $A,B,C,$ and $D$ contributing to the nonperturbative ring potential $\mathcal{V}_{\text{ring}}$. Dashed and solid lines correspond to $\varphi_{1}$ and $\varphi_{2}$, respectively. They are given by the expressions from \eqref{E44}.}\label{fig6}
\end{figure*}
We finally consider the nonperturbative ring potential $\mathcal{V}_{\text{ring}}$. As mentioned in the previous paragraphs, the Lagrangian is written in terms of $\varphi_{1}$ and $\varphi_{2}$, three type of vertices appear in the $\lambda(\varphi^{\star}\varphi)$ model (see Fig. \ref{fig3}). We thus have four different types of ring diagrams:
\begin{itemize}
\item[-] Type A: A ring with $N$ insertions of $\Pi_{2}$ and $N$ propagators $D_{\ell}(\omega_{n},\omega_{1})$ propagators, $\mathcal{V}_{\text{ring}}^{A}$,
\item[-] Type B: A ring with $N$ insertions of $\Pi_{1}$ and $N$ propagators $D_{\ell}(\omega_{n},\omega_{2})$ propagators, $\mathcal{V}_{\text{ring}}^{B}$,
\item[-] Type C: A ring with $r$ insertions of $\Pi_{2}$ and $s$ insertions of $\Pi_{1}$ with $N$ propagators $D_{\ell}(\omega_{n},\omega_{2})$,  $\mathcal{V}_{\text{ring}}^{C}$. Here, $r\geq 1$ and $r+s=N$.
\item[-] Type D: A ring with $r$ insertions of $\Pi_{1}$ and $s$ insertions of $\Pi_{2}$ with $N$ propagators $D_{\ell}(\omega_{n},\omega_{1})$,  $\mathcal{V}_{\text{ring}}^{D}$. Similar to the previous case, $r\geq 1$ and $r+s=N$.
\end{itemize}
Here, $\Pi_{i}(T,\Omega,m_{i})$ and $D_{\ell}(\omega_{n},\omega_{i}), i=1,2$ are defined in \eqref{E16} and \eqref{E17}, respectively. In Fig. \ref{fig6}, these different types of ring potentials are demonstrated. The full contribution of the ring potential is given by
\begin{eqnarray}\label{E43}
\mathcal{V}_{\text{ring}}=\sum_{I=\{A,\cdots,D\}}\mathcal{V}_{\text{ring}}^{I}.
\end{eqnarray}
Following standard field theoretical method, it is possible to determine the combinatorial factors leading to the standard form of the ring potential \cite{kapusta-book}. In Appendix \ref{appD}, we outline the derivation of $\mathcal{V}_{\text{ring}}^{I}, I=A,\cdots,D$. They are given by
\begin{eqnarray}\label{E44}
\mathcal{V}_{\text{ring}}^{\text{A}}&=&-\frac{T}{2}\sum_{n,\ell}\int d\tilde{k}\sum_{N=2}^{\infty}\frac{1}{N}\left(-\Pi_{2}D_{1}\right)^{N},\nonumber\\
\mathcal{V}_{\text{ring}}^{\text{B}}&=&-\frac{T}{2}\sum_{n,\ell}\int d\tilde{k}\sum_{N=2}^{\infty}\frac{1}{N}\left(-\Pi_{1}D_{2}\right)^{N},\nonumber\\
\mathcal{V}_{\text{ring}}^{\text{C}}&=&-\frac{T}{2}\sum_{n,\ell}\int d\tilde{k}\sum_{N=2}^{\infty}\sum_{r=1}^{N}\frac{(N-r)! (r-1)!}{N!}\nonumber\\
&&\times \big[\left(-\Pi_{2}\right)^{r}\left(-\Pi_{1}\right)^{N-r}D_{2}^{N}\big],\nonumber\\
\mathcal{V}_{\text{ring}}^{\text{D}}&=&-\frac{T}{2}\sum_{n,\ell}\int d\tilde{k}\sum_{N=2}^{\infty}\sum_{r=1}^{N}\frac{(N-r)! (r-1)!}{ N!}\nonumber\\
&&\times \big[\left(-\Pi_{1}\right)^{r}\left(-\Pi_{2}\right)^{N-r}D_{1}^{N}\big].
\end{eqnarray}
Here, the notation $D_{i}\equiv D_{\ell}(\omega_{n},\omega_{i})$ is used. To evaluate $\mathcal{V}_{\text{ring}}^{A}$ and $\mathcal{V}_{\text{ring}}^{B}$, we introduce a simplifying notation
\begin{eqnarray}\label{E45}
\mathcal{V}_{\text{ring}}^{(i,j)}=-\frac{T}{2}\sum_{n,\ell}\int d\tilde{k}\sum_{N=2}^{\infty}\frac{1}{N}\left(-\Pi_{i}D_{j}\right)^{N}.
\end{eqnarray}
Here, $(i,j)=(2,1)$ and $(i,j)=(1,2)$ correspond to $\mathcal{V}_{\text{ring}}^{A}$ and $\mathcal{V}_{\text{ring}}^{B}$, respectively. Plugging $D_{j}$ from \eqref{E17} into \eqref{E45} and focusing on $n=0$ as well as $\ell\neq 0$ contributions in the summation over $n$ and $\ell$, we arrive first at
\begin{eqnarray}\label{E46}
\mathcal{V}_{\text{ring}}^{(i,j)}=T\sum_{\ell=1}^{\infty}\int d\tilde{k}\sum_{N=2}^{\infty}\frac{(-1)^{N+1}}{N} (u_{j}^{2})^{-N}\left(\Pi_{i}\right)^{N},\nonumber\\
\end{eqnarray}
with $u_{j}^{2}\equiv \bs{k}_{\perp}^{2}+k_{z}^{2}+m_{j}^{2}-\ell^{2}\Omega^{2}$.
Plugging then
\begin{eqnarray}\label{E47}
(u_{j}^{2})^{-N}=\frac{1}{\Gamma(N)}\int_{0}^{\infty}dt~t^{N-1}e^{-m_{j}^{2}t}e^{-(\bs{k}_{\perp}^{2}+k_{z}^{2})t}~e^{\ell^{2}\Omega^{2}t}, \nonumber\\
\end{eqnarray}
into \eqref{E46}, the integration over $k_{\perp}$ and $k_{z}$ can be carried out by making used of \eqref{appA9}.
To limit the summation over $\ell$ from below, we use the fact that the summand is even in $\ell$.
To perform the integration over $k_{\perp}$ and $k_{z}$, we use the Mellin transformation of $(u_{j}^{2})^{-N}$,
\begin{eqnarray}\label{E48}
(u_{j}^{2})^{-N}=\frac{1}{\Gamma(N)}\int_{0}^{\infty}dt~t^{N-1}e^{-m_{j}^{2}t}e^{-(\bs{k}_{\perp}^{2}+k_{z}^{2})t}~e^{\ell^{2}\Omega^{2}t},  \nonumber\\
\end{eqnarray}
and \eqref{appA9} to arrive first at
\begin{eqnarray}\label{E49}
\mathcal{V}_{\text{ring}}^{(i,j)}=\frac{T}{8\pi^{3/2}}\sum_{N=2}^{\infty}\frac{(-1)^{N+1}\Pi_{i}^{N}}{N\Gamma(N)}\int_{0}^{\infty}dt t^{N-5/2}e^{-m_{j}^{2}t}I(\Omega), \nonumber\\
\end{eqnarray}
where
\begin{eqnarray}\label{E50}
I(\Omega)\equiv \sum_{\ell=1}^{\infty}e^{\ell^{2}\Omega^{2}t}.
\end{eqnarray}
To evaluate the summation over $\ell$, we expand $e^{\ell^{2}\Omega^{2}t}$ in a Taylor expansion and obtain
\begin{eqnarray}\label{E51}
I(\Omega)=\sum_{r=0}^{\infty}\frac{(\Omega^{2} t)^{r}}{r!}\zeta(-2r),
\end{eqnarray}
with $\sum_{\ell=1}^{\infty}\ell^{2r}=\zeta(-2r)$ and $\zeta(z)$ the Riemann $\zeta$-function. Since for $r\in\mathbb{N}$, we have  $\zeta(-2r)=0$, the only nonvanishing contribution to the summation over $r$ arises from $r=0$. We thus use $\zeta(0)=-\frac{1}{2}$ to arrive at
\begin{eqnarray}\label{E52}
I(\Omega)=-\frac{1}{2}.
\end{eqnarray}
Plugging this result into \eqref{E49}, using
\begin{eqnarray}\label{E53}
\int_{0}^{\infty}dt t^{N-5/2}e^{-m_{j}^{2}t}=\left(m_{j}^{2}\right)^{-j+3/2}\Gamma\left(j-3/2\right), \nonumber\\
\end{eqnarray}
and performing the summation over $N$, we arrive at
\begin{eqnarray}\label{E54}
\mathcal{V}_{\text{ring}}^{(i,j)}=\frac{T}{24\pi}\left(2\left(m_{j}^{2}+\Pi_{i}\right)^{3/2}-2m_{j}^{3}-3m_{j}\Pi_{i}\right). \nonumber\\
\end{eqnarray}
We arrive eventually at
\begin{eqnarray}\label{E55}
\mathcal{V}_{\text{ring}}^{A}=\mathcal{V}_{\text{ring}}^{(2,1)},\qquad
\mathcal{V}_{\text{ring}}^{B}=\mathcal{V}_{\text{ring}}^{(1,2)}.
\end{eqnarray}
To evaluate $\mathcal{V}_{\text{ring}}^{C}$ and $\mathcal{V}_{\text{ring}}^{D}$, we introduce
\begin{eqnarray}\label{E56}
V_{\text{ring}}^{(i,j)}&=&-\frac{T}{2}\sum_{n,\ell}\int d\tilde{k}\sum_{N=2}^{\infty}\sum_{r=1}^{N}\frac{(-1)^{N}(N-r)! (r-1)!}{N!}\nonumber\\
&&\times \Pi_{i}^{r}\Pi_{j}^{N-r}D_{i}^{N}.\nonumber\\
\end{eqnarray}
Here, $(i,j)=(2,1)$ corresponds to $\mathcal{V}_{\text{ring}}^{C}$ and $(i,j)=(1,2)$ to $\mathcal{V}_{\text{ring}}^{D}$. Plugging $D_{i}$ from \eqref{E17} into \eqref{E56} and focusing on $n=0$ and $\ell\neq 0$ contributions in the summation over $n$ and $\ell$, we obtain
\begin{eqnarray}\label{E57}
V_{\text{ring}}^{(i,j)}&=&
T\sum_{\ell=1}^{\infty}\int d\tilde{k}\sum_{N=2}^{\infty}\sum_{r=1}^{N}
\frac{(-1)^{N}(N-r)! (r-1)!}{N!}\nonumber\\
&&\times \Pi_{i}^{r}\Pi_{j}^{N-r}(u_{i}^{2})^{-N},
\end{eqnarray}
where $u_{j}^{2}$ is defined below \eqref{E46}. Following, at this stage, the same steps as described in previous paragraph, we arrive first at
\begin{eqnarray}\label{E58}
V_{\text{ring}}^{(i,j)}&=&\frac{m_{i}^{3}T}{16\pi^{3/2}}\sum_{N=2}^{\infty}\sum_{r=1}^{N}\frac{(-1)^{N}}{N!}\frac{(N-r)! (r-1)!}{\Gamma(N)}\nonumber\\
&&\times \Gamma\left(N-3/2\right)\Pi_{i}^{r}\Pi_{j}^{N-r}(m_{i}^{2})^{-N}.
\end{eqnarray}
To perform the summation over $N$ and $r$, we use the relation
\begin{eqnarray}\label{E59}
\sum_{N=2}^{\infty}\sum_{r=1}^{N}f(N,r)=\sum_{N=2}^{\infty}f(N,N)+\sum_{r=1}^{\infty}\sum_{N=r+1}^{\infty}f(N,r). \nonumber\\
\end{eqnarray}
We thus obtain
\begin{eqnarray}\label{E60}
V_{\text{ring}}^{(i,j)}=V^{(i)}+V^{(i,j)},
\end{eqnarray}
with
\begin{eqnarray}\label{E61}
V^{(i)}&\equiv& \frac{m_{i}^{3}T}{16\pi^{3/2}}\sum_{N=2}^{\infty}\frac{(-1)^{N}}{N}\frac{\Gamma(N-3/2)\Pi_{i}^{N}(m_{i}^{2})^{-N}}{\Gamma(N)}\nonumber\\
V^{(i,j)}&\equiv&\frac{m_{i}^{3}T}{16\pi^{3/2}}
\sum_{r=1}^{\infty}\sum_{N=r+1}^{\infty}
\frac{(-1)^{N}}{N!}\frac{(N-r)! (r-1)!}{\Gamma(N)}\nonumber\\
&&\times \Gamma(N-3/2)\Pi_{i}^{r}\Pi_{j}^{N-r}(m_{i}^{2})^{-N}.
\end{eqnarray}
For $V^{(i)}$, the summation over $N$ can be carried out and yields
\begin{eqnarray}\label{E62}
V^{(i)}=\frac{T}{24}\left(2\left(m_{i}^{2}+\Pi_{i}\right)^{3/2}-2m_{i}^{3}-3m_{i}\Pi_{i}\right). \nonumber\\
\end{eqnarray}
As concerns $V^{(i,j)}$, we perform the summation over $N$ and arrive at
\begin{eqnarray}\label{E63}
V^{(i,j)}&=&
\sum_{r=1}^{\infty}\frac{ (-1)^{r+1}}{r}\frac{\Gamma \left(r-1/2\right)}{\Gamma (r+2)}
\Pi _{i}^r \Pi _{j}\left(m_{i}^{2}\right)^{-r-1}\nonumber\\
&&\times _3F_2\left((1,2,r-1/2);(r+1,r+2);-\frac{\Pi_{j}}{m_{i}^{2}}\right), \nonumber\\
\end{eqnarray}
where $_pF_q\left(\bs{a};\bs{b};z\right)$ is the generalized hypergeometric function having the following series expansion
\begin{eqnarray}\label{E64}
_pF_q\left(\bs{a};\bs{b};z\right)=\sum_{k=0}^{\infty}\frac{(a_{1})_{k}\cdots(a_{p})_{k}}{(b_{1})_{k}\cdots (b_{q})_{k}}\frac{z^{k}}{k!}.
\end{eqnarray}
Here, $\bs{a}=(a_{1},\cdots,a_{p})$, $\bs{b}=(b_{1},\cdots,b_{q})$ are vectors with $p$ and $q$ components. Moreover, $(a_{i})_{k}\equiv \Gamma\left(a_{i}+k\right)/\Gamma\left(a_{i}\right)$ is the Pochhammer symbol. For our purposes, it is sufficient to focus on the contribution at $r = 1$ in \eqref{E63}.
\begin{eqnarray}\label{E65}
V^{(i,j)}|_{r=1}=\frac{T}{24\pi}\left(2\left(m_{i}^{2}+\Pi_{j}\right)^{3/2}-2m_{i}^{3}-3m_{i}\Pi_{j}\right)\frac{\Pi_{i}}{\Pi_{j}}.\nonumber\\
\end{eqnarray}
Having in mind that the one-loop contribution to the self-energy $\Pi_{i}$, which is determined in Sec. \ref{sec3B} is of order ${O}(\lambda)$, the contributions corresponding to $r\geq 2$ are of order $\mathcal{O}(\lambda^{2})$ and can be neglected at this stage. We thus have
\begin{eqnarray}\label{E66}
\mathcal{V}_{\text{ring}}^{C}&=& \mathcal{V}_{\text{ring}}^{(2,1)}=V^{(2)}+V^{(2,1)}|_{r=1}+\mathcal{O}(\lambda^{2}), \nonumber\\
\mathcal{V}_{\text{ring}}^{D}&=& \mathcal{V}_{\text{ring}}^{(1,2)}=V^{(1)}+V^{(1,2)}|_{r=1}+\mathcal{O}(\lambda^{2}).\nonumber\\
\end{eqnarray}
The final result for $\mathcal{V}_{\text{ring}}$ is given by plugging $\mathcal{V}_{\text{ring}}^{I}, I=A,\cdots,D$ from \eqref{E55} and \eqref{E65} into \eqref{E43},
\begin{eqnarray}\label{E67}
\mathcal{V}_{\text{ring}}&=&\frac{T}{24\pi}\bigg\{\left(2\left(m_{1}^{2}+\Pi_{2}\right)^{3/2}-2m_{1}^{3}-3m_{1}\Pi_{2}\right)\nonumber\\
&&+\left(2\left(m_{2}^{2}+\Pi_{1}\right)^{3/2}-2m_{2}^{3}-3m_{2}\Pi_{1}\right)\nonumber\\
&&+\left(2\left(m_{2}^{2}+\Pi_{2}\right)^{3/2}-2m_{2}^{3}-3m_{2}\Pi_{2}\right)\nonumber\\
&&+\left(2\left(m_{1}^{2}+\Pi_{1}\right)^{3/2}-2m_{1}^{3}-3m_{1}\Pi_{1}\right)\nonumber\\
&&+\left(2\left(m_{2}^{2}+\Pi_{1}\right)^{3/2}-2m_{2}^{3}-3m_{2}\Pi_{1}\right)\frac{\Pi_{2}}{\Pi_{1}}\nonumber\\
&&+\left(2\left(m_{1}^{2}+\Pi_{2}\right)^{3/2}-2m_{1}^{3}-3m_{1}\Pi_{2}\right)\frac{\Pi_{1}}{\Pi_{2}}\bigg\}\nonumber\\
&&+\mathcal{O}(\lambda^{2}).
\end{eqnarray}
Focusing only on the first perturbative correction to $\Pi_{i}$ and using $\Pi_{i}^{\text{mat}}, i=1,2$ from \eqref{E28}, the above results is simplified as
\begin{eqnarray}\label{E68}
\mathcal{V}_{\text{ring}}\approx
\frac{T}{8\pi}\sum_{i=1}^{2}\left(2\left(m_{i}^{2}+\Pi^{\text{mat}}\right)^{3/2}-2m_{i}^{3}-3m_{i}\Pi^{\text{mat}}\right), \nonumber\\
\end{eqnarray}
where $\Pi^{\text{mat}}\equiv \Pi_{1}^{\text{mat}}=\Pi_{2}^{\text{mat}}=\dfrac{\lambda T^{3}\zeta(3)}{2\pi^{2}\Omega}$.
\subsection{Summary of analytical results in Sec. \ref{sec3}}\label{sec3F}
In this section, we summarize the main findings. According to these results, the total thermodynamic potential of a rigidly rotating Bose gas, $\mathcal{V}_{\text{tot}}$, including the classical potential $\mathcal{V}_{\text{cl}}$ from \eqref{N32} with $c^{2}$ replaced with $a^{2}$, the vacuum potential \eqref{N33}, the thermal part \eqref{N34}, and the ring potential \eqref{E43} is given by
\begin{eqnarray}\label{E69}
\mathcal{V}_{\text{tot}}=\mathcal{V}_{\text{cl}}+\mathcal{V}_{\text{vac}}+\mathcal{V}_{T}+\mathcal{V}_{\text{ring}},
\end{eqnarray}
with
\begin{eqnarray}\label{E70}
\mathcal{V}_{\text{cl}}&=&-\frac{a^{2}v^{2}}{2}+\frac{\lambda v^{4}}{4},\nonumber\\
\mathcal{V}_{\text{vac}}&\approx&
-\frac{m_{0}^{2}v^{2}}{4}+\frac{3c^{2}\lambda v^{2}}{8\pi^{2}}+\frac{5m_{0}^{2}\lambda v^{2}}{16\pi^{2}}-\frac{15\lambda^{2}v^{4}}{64\pi^{2}}+\frac{m_{0}^{2}\lambda v^{4}}{8a^{2}},
\nonumber\\
\mathcal{V}_{T}&\approx&
-\frac{2T^{5}\zeta(5)}{\pi^{2}\Omega}+\frac{\lambda T^{3}v^{2}\zeta(3)}{\pi^{2}\Omega}-\frac{c^{2}T^{3}\zeta(3)}{2\pi^{2}\Omega},\nonumber\\
\mathcal{V}_{\text{ring}}&\approx&
+\frac{T}{8\pi}\sum_{i=1}^{2}\left(2\left(m_{i}^{2}+\Pi^{\text{mat}}\right)^{3/2}-2m_{i}^{3}-3m_{i}\Pi^{\text{mat}}\right). \nonumber\\
\end{eqnarray}
Here, $a^{2}=c^{2}+m_{0}^{2}$, $m_{1}^{2}(v)=3\lambda v^{2}-c^{2}$ and $m_{2}^{2}(v)=\lambda v^{2}-c^{2}$, and
 $\Pi^{\text{mat}}=\lambda T^{3}\zeta(3)/2\pi^{2}\Omega$. We notice that the logarithmic terms appearing in $\mathcal{V}_{\text{vac}}$ from \eqref{E42} are skipped in \eqref{E70}.
\par
In the next section, we study the effect of rotation on the formation of condensate and the critical temperature of the global $\mathrm{U}(1)$ phase transition. To this purpose, we compare our results with the results arising from the full thermodynamic potential of a nonrotating Bose gas. According to \cite{kapusta-book}, it is given by\footnote{Subscripts $(0)$ correspond to $\Omega=0$. }
\begin{eqnarray}\label{E71}
\mathcal{V}_{\text{tot}}^{(0)}=\mathcal{V}_{\text{cl}}+\mathcal{V}_{\text{vac}}+\mathcal{V}_{T}^{(0)}+\mathcal{V}_{\text{ring}}^{(0)},
\end{eqnarray}
where $\mathcal{V}_{\text{cl}}$ and $\mathcal{V}_{\text{vac}}$ are given in \eqref{E70}, while
$\mathcal{V}_{T}^{(0)}$ and $\mathcal{V}_{\text{ring}}^{(0)}$ read
\begin{eqnarray}\label{E72}
\mathcal{V}_{T}^{(0)}\approx-\frac{\pi^{2}T^{4}}{45}+\frac{\lambda T^{2}v^{2}}{6}-\frac{c^{2}T^{2}}{12},
\end{eqnarray}
and
\begin{eqnarray}\label{E73}
\mathcal{V}_{\text{ring}}^{(0)}\approx
-\frac{T}{4\pi}\sum_{i=1}^{2}\left(2\left(m_{i}^{2}+\Pi_{0}^{\text{mat}}\right)^{3/2}-2m_{i}^{3}-3m_{i}\Pi_{0}^{\text{mat}}\right), \nonumber\\
\end{eqnarray}
with the one-loop self-energy correction $\Pi_{0}^{\text{mat}}=\lambda T^{2}/3$ \cite{kapusta-book} and $m_{i}^{2}, i=1,2$ given as above.
\section{Numerical results}\label{sec4}
\setcounter{equation}{0}
\begin{figure}
\includegraphics[width=8cm, height=5.5cm]{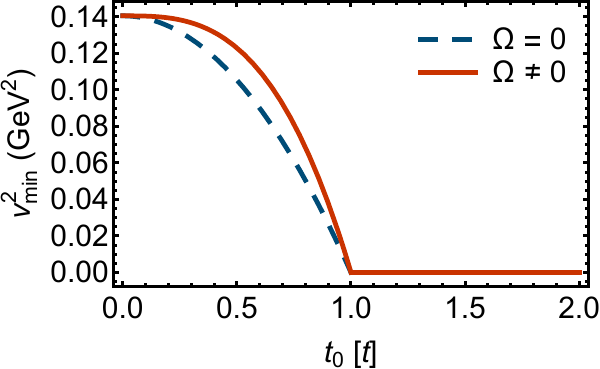}
\caption{(color online). The $t_{0} [t]$ dependence of $v_{\text{min}}^{2}(T)$ and $v_{\text{min}}^{2}(T,\Omega)$ for nonrotating ($\Omega=0$) and rotating ($\Omega\neq 0$) Bose gas [see \eqref{E6} and \eqref{D2}]. For $\Omega=0$ and $\Omega\neq 0$, the reduced temperature  $t_{0}$ or $t$ is defined by $t_{0}=T/T_{c}^{(0)}$ and $t=T/T_{c}$, respectively.}\label{fig7}
\end{figure}
In this section, we explore the effect of rotation on different quantities related to the spontaneous breaking of global $\mathrm{U}(1)$ symmetry. To this purpose, we consider different parts of $\mathcal{V}_{\text{tot}}$ from \eqref{E69}.
\par
In Sec. \ref{sec3A}, we derived the minimum of the potential $\mathcal{V}_{a}$ including $\mathcal{V}_{\text{cl}}$ and $\mathcal{V}_{T}$. We arrived at $v_{\text{min}}^{2}(T,\Omega)$ from \eqref{E6}. Replacing $\mathcal{V}_{T}$ with $\mathcal{V}_{T}^{(0)}$ from \eqref{E72} for a nonrotating Bose gas and following the same steps leading from \eqref{E4} to \eqref{E6}, we arrive at the critical temperature
\begin{eqnarray}\label{D1}
T_{c}^{(0)}=\left(\frac{3a^{2}}{\lambda}\right)^{1/2},
\end{eqnarray}
and the $T$ dependent minima
\begin{eqnarray}\label{D2}
v_{\text{min}}^{2}(T)=
\begin{cases}
\dfrac{a^{2}}{\lambda}\left(1-t_{0}^{2}\right),&t_{0}< 1,\\
0,&t_{0}\geq 1,
\end{cases}
\end{eqnarray}
with the reduced temperature $t_{0}= T/T_{c}^{(0)}$ and $T_{c}^{(0)}$ from \eqref{D1}. In Fig. \ref{fig7}, $v_{\text{min}}^{2}$ is plotted for $\Omega=0$ [see \eqref{D2}] and $\Omega\neq 0$ [see \eqref{E6}] as function of the corresponding reduced temperature $t_{0}$ and $t$. The difference between these two plots arises mainly from different exponents of the corresponding reduced temperatures $t_{0}$ and $t$ in \eqref{D2} and \eqref{E6}. The reason of this difference lies in different results for the high-temperature expansion of $\mathcal{V}_{T}^{(0)}$ for $\Omega=0$ [see \eqref{E72}] and $\mathcal{V}_{T}$ for $\Omega\neq 0$ [see \eqref{E70}].
\begin{figure}
\includegraphics[width=8cm, height=5.5cm]{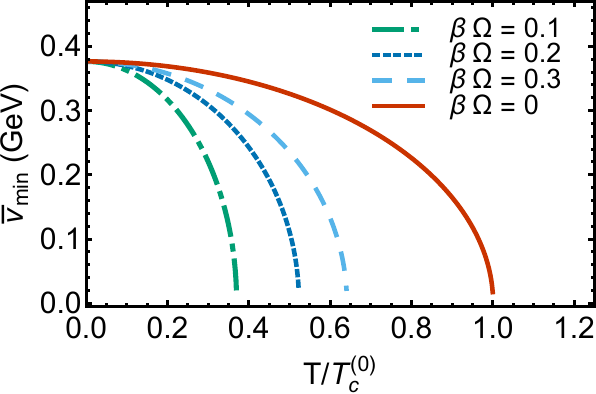}
\caption{(color online).   The $T/T_{c}^{(0)}$ dependence of $\bar{v}_{\text{min}}$ is plotted for $\beta\Omega=0,0.1,0.2, 0.3$. For $\Omega\neq 0$ and $\Omega=0$, $\bar{v}_{\text{min}}(T)$ arises by solving the gap equation \eqref{D3} and \eqref{D4}, respectively. The temperature $T$ is rescaled with $T_{c}^{(0)}=0.681$ GeV, the $\Omega$ independent critical temperature of a nonrotating Bose gas. It turns out that $T_{c}<T_{c}^{(0)}$ and for $\beta\Omega\neq 0$, $T_{c}$ increases by increasing $\beta\Omega$.}\label{fig8}
\end{figure}
\begin{figure}
\includegraphics[width=8cm, height=5.5cm]{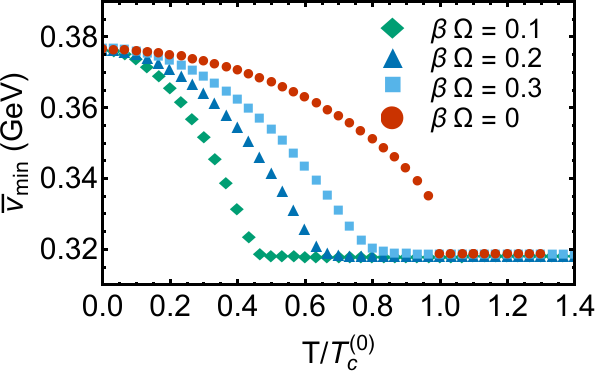}
\caption{(color online).  The $T/T_{c}^{(0)}$ dependence of $\bar{v}_{\text{min}}$ is plotted for $\beta\Omega=0,0.1,0.2,0.3$.  For $\Omega\neq 0$ and $\Omega=0$, $\bar{v}_{\text{min}}(T)$ arises by solving the gap equation \eqref{D5} and \eqref{D6}, respectively. Here, $\bar{v}_{\star}=0.319$ GeV and $T_{\star}^{(0)}=0.300$ GeV. It turns out that $T_{\star}<T_{\star}^{(0)}$ and for $\beta\Omega\neq 0$, $T_{\star}$ increases by increasing $\beta\Omega$.}\label{fig9}
\end{figure}
\par
Let us consider $\mathcal{V}_{\text{tot}}-\mathcal{V}_{\text{ring}}=\mathcal{V}_{\text{cl}}+\mathcal{V}_{\text{vac}}+\mathcal{V}_{T}$ from \eqref{E69}. By minimizing this potential with respect to $v$, and solving the resulting gap equation,
\begin{eqnarray}\label{D3}
\frac{d}{d v}\left(\mathcal{V}_{\text{tot}}-\mathcal{V}_{\text{ring}}\right)\bigg|_{\bar{v}_{\text{min}}}=0,
\end{eqnarray}
it is possible to determine numerically the $T$ dependence the minima, denoted by $\bar{v}_{\text{min}}(T,\Omega)$, for fixed $\Omega$. To this purpose, we use the quantities $a\simeq 0.265$ GeV, $c\simeq 0.225$ GeV, and $\lambda=0.5$ given in \eqref{E7}. In Fig. \ref{fig8}, the $T/T_{c}^{(0)}$ dependence of $\bar{v}_{\text{min}}$ is demonstrated for $\beta\Omega=0.1,0.2,0.3$ (dashed, dotted, and dotted-dashed curves).
The results are then compared with the corresponding minima for a nonrotating Bose gas (red solid curve). The latter is determined by minimizing the combination $\mathcal{V}_{\text{tot}}^{(0)}-\mathcal{V}_{\text{ring}}^{(0)}$, according to
\begin{eqnarray}\label{D4}
\frac{d}{d v}\left(\mathcal{V}_{\text{tot}}^{(0)}-\mathcal{V}_{\text{ring}}^{(0)}\right)\bigg|_{\bar{v}_{\text{min}}}=0,
\end{eqnarray}
with $\mathcal{V}_{\text{tot}}^{(0)}$ from \eqref{E71}. In both cases, $T_{c}^{(0)}\simeq 0.681$ GeV is the critical temperature of the spontaneous $\mathrm{U}(1)$ symmetry breaking in a nonrotating Bose gas.\footnote{The critical temperature is the temperature at which the condensate $\bar{v}_{\text{min}}$ vanishes.}
\\
These results indicate that rotation lowers the critical temperature of the phase transition. However, as shown in Fig. \ref{fig8}, $T_c$ increases with increasing $\Omega$. It is also important to note that this same trend is observed in a noninteracting Bose gas under rigid rotation \cite{siri2024b}.
\par
To answer the question whether the transition is continuous or discontinuous, we have to explore the shape of the potential, the value of its first and second order derivatives at temperatures below and above the critical temperature, $T_{c}$. Using the numerical values for the set of free parameters $a,c$, and $\lambda$ as mentioned above, the transitions turns out to be continuous not only for $\Omega=0$ but also for $\Omega\neq 0$.
\par
To explore the effect of the ring potential on the temperature dependence of the condensate $\bar{v}_{\text{min}}$, we solved numerically the gap equation
\begin{eqnarray}\label{D5}
\frac{d\mathcal{V}_{\text{tot}}}{d v}\bigg|_{\bar{v}_{\text{min}}}=0,
\end{eqnarray}
and
\begin{eqnarray}\label{D6}
\frac{d\mathcal{V}_{\text{tot}}^{(0)}}{d v}\bigg|_{\bar{v}_{\text{min}}}=0,
\end{eqnarray}
for a rotating and a nonrotating Bose gas, respectively. The corresponding results are demonstrated in Fig. \ref{fig9}.
\begin{figure}
\includegraphics[width=8cm, height=5.5cm]{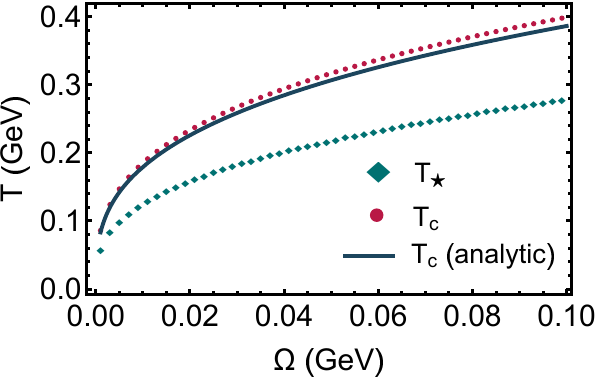}
\caption{(color online).  The $\Omega$ dependence of the transition temperatures is plotted. The blue solid line is the transition temperature $T_{c}\propto \Omega^{1/3}$ from \eqref{E5}. It arises from $\mathcal{V}_{\text{cl}}+\mathcal{V}_{T}$, as described in Sec. \ref{sec3A}. Red dots correspond to the critical temperatures $T_{c}$, arising from $\mathcal{V}_{\text{tot}}-\mathcal{V}_{\text{ring}}$. Green diamonds denote $T_{\star}$, arising from $\mathcal{V}_{\text{tot}}$.}\label{fig10}
\end{figure}
Because of the specific form of the ring potentials $\mathcal{V}_{\text{ring}}$ and $\mathcal{V}_{\text{ring}}^{(0)}$ from \eqref{E70} and \eqref{E73}, including in particular $(m_{i}^{2}+\Pi^{\text{mat}})^{3/2}$, there is a certain value of $v$ below which the potential is undefined (imaginary). Let us denote this value by $\bar{v}_{\star}$. In both rotating and nonrotating cases $v_{\star}\simeq 0.319$ GeV. As it is shown in Fig. \ref{fig9}, the minima decrease with increasing temperature and converge towards $\bar{v}_{\star}$. Let us denote the temperature at which $\bar{v}_{\text{min}}=v_{\star}$ with $T_{\star}$ for $\Omega\neq 0$ and $T_{\star}^{(0)}$ for $\Omega=0$. For $\Omega=0$, $T_{\star}^{(0)}\simeq 0.300$ GeV, and as it is shown in Fig. \ref{fig9}, the transition to $v_{\star}$ is discontinuous (red circles). For $\Omega\neq 0$, however, $T_{\star}<T_{\star}^{(0)}$ and increases with increasing $\beta\Omega$, similar to the results presented in Fig. \ref{fig8}. Moreover, in contrast to the case of $\Omega=0$, the transition to $\bar{v}_{\star}$ for all values of $\beta\Omega\neq 0$ is continuous.
\par
In Fig. \ref{fig10}, the phase diagram $T_{c}$-$\Omega$ is plotted for two cases: The blue solid curve demonstrates $T_{c}$ from \eqref{E5} arising from $\mathcal{V}_{\text{cl}}+\mathcal{V}_{T}$.
Red dots denote the $\Omega$ dependence of $T_{c}$ arising from the potential $\mathcal{V}_{\text{tot}}-\mathcal{V}_{\text{ring}}$. A comparison between these data reveals the effect of $\mathcal{V}_{\text{vac}}$ in increasing $T_{c}$. Apart from the $\Omega$ dependence of $T_{c}$, the $\Omega$ dependence of $T_{\star}$ is demonstrated in Fig. \ref{fig10}. It arises by adding the ring contribution to $\mathcal{V}_{\text{cl}}+\mathcal{V}_{T}+\mathcal{V}_{\text{vac}}$, as described above. According to the results demonstrated in Fig. \ref{fig10}, considering $\mathcal{V}_{\text{ring}}$ decreases $T_{c}$. But, similar to $T_{c}$, $T_{\star}$ also increases with increasing $\Omega$. It should be emphasized that the transition shown in Fig. \ref{fig8} is a crossover, since $\bar{v}_{\star}\neq 0$.
\par
\begin{figure*}
\includegraphics[width=8cm, height=5.5cm]{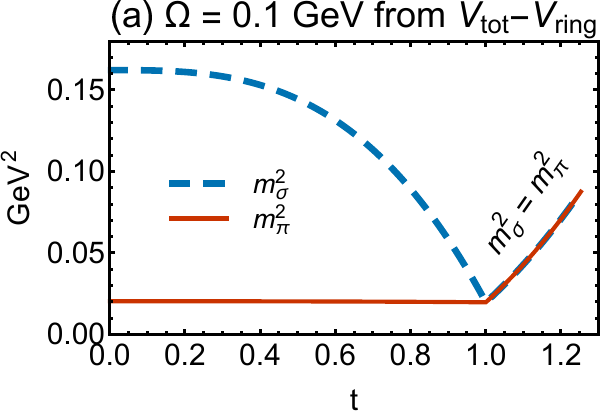}
\includegraphics[width=8cm, height=5.5cm]{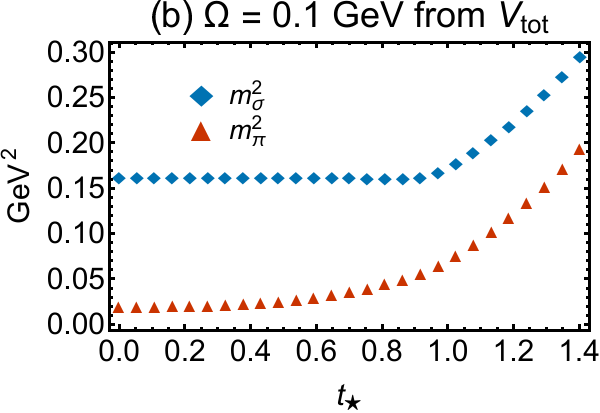}
\caption{(color online). (Panel a) The $t=T/T_{c}$ dependence of $m_{\sigma}^{2}(\bar{v}_{\text{min}})$ and $m_{\pi}^{2}(\bar{v}_{\text{min}})$ from \eqref{D7} is plotted for $\Omega=0.1$ GeV. The data of $\bar{v}_{\text{min}}$ arise by solving the gap equation \eqref{D3} corresponding to $\mathcal{V}_{\text{tot}}-\mathcal{V}_{\text{ring}}$. The critical temperature $T_{c}$ for $\Omega=0.1$ GeV is $T_{c}\sim 0.399$ GeV. As expected from the case of nonrotating Bose gas, in the symmetry-restored phase at $t\geq 1$, $m_{\sigma}^{2}=m_{\pi}^{2}$.  (Panel b) The $t_{\star}$ dependence of $m_{\sigma}^{2}(\bar{v}_{\star})$ and $m_{\pi}^{2}(\bar{v}_{\star})$ from \eqref{D7} is plotted for $\Omega=0.1$ GeV. The data of $\bar{v}_{\text{min}}$ arise by solving the gap equation \eqref{D5}, corresponding to $\mathcal{V}_{\text{tot}}$ which includes the nonperturbative ring potential. According to Fig. \ref{fig10}, for $\Omega=0.1$ GeV, we have $T_{\star}\sim 0.278$ GeV. At $t\geq 1$, $m_{\sigma}^{2}-m_{\pi}^{2}=2\lambda\bar{v}_{\star}^{2}$, with $\bar{v}_{\star}\simeq 0.319$ GeV from Fig. \ref{fig9} and $\lambda=0.5$.}\label{fig11}
\end{figure*}
In Sec. \ref{sec3B}, the masses $m_{i}^{2}, i=1,2$ including the one-loop correction are determined [see \eqref{E29}]. Identifying $m_{1}^{2}$ with $m_{\sigma}^{2}$ and $m_{2}^{2}$ with $m_{\pi}^{2}$, we arrive at
\begin{eqnarray}\label{D7}
m_{\sigma}^{2}(v)&=&3\lambda v^{2}-c^{2}+a^{2} t^{3},\nonumber\\
m_{\pi}^{2}(v)&=&\lambda v^{2}-c^{2}+a^{2} t^{3}.
\end{eqnarray}
Using the data for $\bar{v}_{\text{min}}^{2}$ arising from the solution of the gap equation \eqref{D3} and \eqref{D5}, and evaluating $m_{\sigma}^{2}(v^{2})$ and $m_{\pi}^{2}(v^{2})$ from \eqref{D7} at $\bar{v}^{2}_{\text{min}}$ for a fixed $\beta\Omega$, the $t=T/T_{c}$ dependence of $m_{\sigma}^{2}$ and $m_{\pi}^{2}$ is determined.
In Fig. \ref{fig11}(a), the dependence of $m_{\sigma}^{2}(\bar{v}_{\text{min}}^{2})$ and $m_{\pi}^{2}(\bar{v}_{\text{min}}^{2})$ with $v_{\text{min}}$ arising from \eqref{D3} on the reduced temperature $t=T/T_{c}$ is plotted for fixed $\beta\Omega=0.1$.
Here, the contribution of the ring potential is not taken into account. Hence, a continuous phase transition occurs with the critical temperature $T_{c}\sim 0.399$ GeV for $\Omega=0.1$ GeV. In contrast, in Fig. \ref{fig11}(b), $m_{\sigma}^{2}$ and $m_{\pi}^{2}$ are determined by plugging the data of $\bar{v}_{\text{min}}$ arising from \eqref{D5}, with $\mathcal{V}_{\text{tot}}$ including the ring potential. Hence, the difference between the plots demonstrated in Figs. \ref{fig11}(a) and \ref{fig11}(b) arises from the contribution of the nonperturbative ring potential.
As we have mentioned above, when the ring potential is taken into account, the data demonstrated in Fig. \ref{fig9} do not describe a true transition, since $\bar{v}_{\star}$ is not zero. The reduced temperature in Fig. \ref{fig11}(b) is thus defined by $t_{\star}\equiv T/T_{\star}$, where, according to the data presented in Fig. \ref{fig10} $T_{\star}\sim 0.278$ GeV for $\Omega=0.1$.
\par
Let us compare the results demonstrated in Fig. \ref{fig11}(a) with that in Fig. \ref{fig5}. In both cases, before the phase transition at $t<1$, $m_{\sigma}^{2}$ decreases with increasing $t$. Moreover, whereas in Fig. \ref{fig5}, $m_{\pi}^{2}$ remains constant, it slightly decreases once the $\mathcal{V}_{\text{vac}}$ contribution is taken into account. After the transition, at $t\geq 1$, $m_{\sigma}^{2}$ becomes equal to $m_{\pi}^{2}$ and they both increase with increasing $t$. It is straightforward to verify this statement using equation \eqref{D7}. Given that, in this case, the minima of the potential at $t \geq 1$ are zero, it follows that both masses are equal, specifically $m_{\sigma}^{2}(0) = m_{\pi}^{2}(0)$, once we substitute $\bar{v}_{\text{min}} = 0$ into \eqref{D7}.
\par
This behavior is expected from the case of $\Omega=0$ and in the framework of fermionic Nambu-Jona--Lasinio (NJL) model: As noted in \cite{quack1995}, in the symmetry-broken phase, $m_{\sigma}^{2} > m_{\pi}^{2}$. As the transition temperature is approached, $m_{\sigma}^{2}$ decreases, and at a certain dissociation temperature $T_{\text{diss}}$, the masses $m_{\sigma}$ and $m_{\pi}$ become degenerate. This temperature is characterized by
\begin{eqnarray}\label{D8}
m_{\sigma}(T_{\text{diss}})=2m_{\pi}(T_{\text{diss}}).
\end{eqnarray}
As it is described in \cite{quack1995}, $\sigma$ mesons dissociates into two pions because of the appearance of an $s$-channel pole in the scattering amplitude $\pi+\pi\to \pi+\pi$. In this process a $\sigma$ meson is coupled to two pions via a quark triangle. In the symmetry-restored phase, at $t\geq 1$, $m_{\sigma}$ becomes equal to $m_{\pi}$. They both increase with increasing $T$ \cite{buballa2012,quack1995}.
\par
In Table \ref{tab1}, the $\sigma$ dissociation temperatures are listed for $\Omega=0,0.1,0.2,0.3$ GeV. The data in the second (third) column correspond to $T_{\text{diss}}$ ($T_{\text{diss}}^{\star}$) for the case when $\bar{v}_{\text{min}}$ is the solution of \eqref{D3} [\eqref{D5}] for $\Omega\neq 0$ and \eqref{D4} [\eqref{D6}] for $\Omega=0$. Comparing $T_{\text{diss}}$ and $T_{\text{diss}}^{\star}$ with $T_{c}$ and $T_{\star}$ shows that $T_{\text{diss}}< T_{c}$ and similarly $T_{\text{diss}}^{\star}<T_{\star}$. The property $T_{\text{diss}}\neq T_{c}$ is because we are working with $m_{\pi}\neq 0$. Let us notice that, as aforementioned, the $\sigma$ dissociation temperature is originally introduced in a fermionic NJL model \cite{quack1995}. In this model, nonvanishing $m_{\pi}$ indicates a nonvanishing quark bare mass $\tilde{m}_{0}$, and choosing $\tilde{m}_{0}\neq 0$ implies a crossover transition characterized by $T_{\text{diss}}\neq T_{c}$. It seems that in the bosonic model studied in the present work, a nonvanishing pion mass leads similarly to $T_{\text{diss}}\neq T_{c}$.
\par
\begin{table}[hbt]
\begin{tabular}{ccccc}
\hline\hline\\
$\Omega$ in GeV&\qquad\qquad&$T_{\text{diss}}$  $[T_{c}]$ in GeV&\qquad\qquad\qquad&
$T_{\text{diss}}^{\star}$ $[T_{\star}]$ in GeV\\ \hline\\
$0$&\qquad\qquad&$0.584~[0.681]$&\qquad\qquad&$0.220~[0.300]$\\
$0.1$&\qquad\qquad&$0.322~[0.399]$&\qquad\qquad&$0.210~[0.278]$\\
$0.2$&\qquad\qquad&$0.418~[0.502]$&\qquad\qquad&$0.271~[0.358]$\\
$0.3$&\qquad\qquad&$0.480~[0.576]$&\qquad\qquad&$0.316~[0.416]$\\
\hline\hline
\end{tabular}
\caption{The $\sigma$ dissociation temperature for a nonrotating gas with $\Omega=0$ and a rotating gas with $\Omega=0.1,0.2,0.3$ GeV is compared with the critical temperature $T_{c}$ and crossover temperature $T_{\star}$. In the second column, the data arise from the solution of the gap equation \eqref{D3} and \eqref{D4}. In the third column, the data arise from the solution of the gap equation \eqref{D5} and \eqref{D6}. In both cases the dissociation temperature is lower than the transition temperatures. }\label{tab1}
\end{table}
\par
The behavior demonstrated in Fig. \ref{fig11}(a) changes once the contribution of the ring potential is taken into account. As it is shown in Fig. \ref{fig11}(b), in the symmetry-broken phase at $t_{\star}<1$, $m_{\sigma}$ decreases slightly with $T$, while $m_{\pi}$ increases with $T$. Moreover, in contrast to the case in which $\mathcal{V}_{\text{ring}}$ is not taken into account, $m_{\sigma}$ and $m_{\pi}$ are not equal at $t\geq 1$. This observation highlights the effect of nonperturbative ring contributions on the relation between $m_{\sigma}$ and $m_{\pi}$, mainly in the symmetry-restored phase.
This behavior is directly related to the fact that the effect illustrated in Fig. \ref{fig9} is a crossover once the ring contribution is considered: Plugging $\bar{v}_{\star}$ into \eqref{D7}, the masses of $\sigma$ and $\pi$ mesons are given by
\begin{eqnarray}\label{D9}
m_{\sigma}^{2}(\bar{v}_{\star})&=&3\lambda \bar{v}_{\star}^{2}-c^{2}+a^{2} t^{3},\nonumber\\
m_{\pi}^{2}(\bar{v}_{\star})&=&\lambda \bar{v}_{\star}^{2}-c^{2}+a^{2} t^{3}.
\end{eqnarray}
Their difference is thus given by $m_{\sigma}^{2}(\bar{v}_{\star})-m_{\pi}^{2}(\bar{v}_{\star})=2\lambda\bar{v}_{\star}^{2}$ and remains constant in $t$. This fact can be observed in Fig. \ref{fig11}(b) at $t_{\star}\geq 1$.
\section{Summary and Conclusions}\label{sec5}
In this paper, we extended the study of the effects of rigid rotation on BE condensation of a free Bose gas in \cite{siri2024b}, to a self-interacting charged Bose gas under rigid rotation. In the first part, we considered the Lagrangian density of a complex scalar field $\varphi$ with mass $m$, in the presence of chemical potential $\mu$ and angular velocity $\Omega$. The interaction was introduced through a $\lambda(\varphi^{\star}\varphi)$ term. This Lagrangian is invariant under global U(1) transformation. To investigate the spontaneous breaking of this symmetry, we chose a fixed minimum with a real component $v$, and evaluated the original Lagrangian around this minimum to derive a classical potential. Then, we applied an appropriate Bessel-Fourier transformation to determine the free propagator of this model, expressed in terms of two masses $m_{1}$ and $m_{2}$, corresponding to the two components of the complex field $\varphi$. These masses depend explicitly on $v,\lambda$, and $m$, and played a crucial role when the spontaneous breaking of U(1) symmetry was considered in a realistic model that includes $\sigma$ and $\pi$ mesons. Using the free boson propagator of this model, we derived the thermodynamic potential of self-interacting Bose gas at finite temperature $T$. This potential consists of a vacuum and a thermal part. Along with the classical potential, this forms the total thermodynamic potential of this model $\mathcal{V}_{\text{tot}}$ from \eqref{N24}. This potential is expressed in terms of the energy dispersion relation $\epsilon_{k}^{\pm}$ from \eqref{N15}, and explicitly depends on $\ell\Omega$. A novel result presented here is that, although $\ell\Omega$ appears to resemble a chemical potential in combination with $\epsilon_{k}^{\pm}$ in $\mathcal{V}_{\text{tot}}$, the chemical potential $\mu$ affects $\epsilon_{k}^{\pm}$ in a nontrivial manner. The effective chemical potential $\mu_{\text{eff}}=\mu+\ell\Omega$ appears solely in a noninteracting Bose gas under rotation (see the special case 1 in Sec. \ref{sec2D} and compare the thermodynamic potential with that appearing in \cite{siri2024b}).
\\
For $\lambda,\mu\neq 0$, we explored two cases $\mu>m$ and $\mu<m$. The former corresponds to the phase where U(1) symmetry is broken, while the latter describes the symmetry-restored phase. By expanding the two branches of the energy dispersion relation around $k\sim 0$ in the symmetry-broken phase, we identified $\epsilon_{k}^{+}$ and $\epsilon_{k}^{-}$ as phonon and roton, with the latter representing a massless Goldstone mode. Upon comparison with analogous results for a nonrotating and self-interacting Bose gas, we found that rigid rotation does not alter the behavior of $\epsilon_{k}^{\pm}$ at $k\sim 0$. This is mainly because rotation appears in terms of $\ell\Omega$ within $\mathcal{V}_{\text{tot}}$, rather than directly affecting $\epsilon_{k}^{\pm}$.
\par
In the second part of this paper, we examined the effect of rigid rotation on the spontaneous breaking of U(1) symmetry in an interacting Bose gas at $\mu=0$ (see Sec. \ref{sec3}). In this case, where $m^{2}<0$, we replaced $m^{2}$ with $-c^{2}$, where $c^{2}>0$. By introducing an additional term to the original Lagrangian, we defined a new mass, $a^{2}=c^{2}+m_{0}^{2}$. We demonstrated that the minimum of the classical potential is nonzero, indicating a spontaneous breaking of U(1) symmetry. We then addressed the question about the position of this minimum, specifically its dependence on $T$ and $\Omega$, after accounting for the thermal part of the effective potential combined with the classical potential. To investigate this, we performed a high-temperature expansion of the thermal part of the potential, utilizing a method originally introduced in \cite{siri2024b}. This approach enabled us to sum over the angular momentum quantum numbers $\ell$ for small values of $\beta\Omega$, allowing us to derive both the critical temperature of the phase transition $T_c$ and the dependencies of the minimum of the potential on $T$ and $\Omega$. At this stage, we have $T_{c}\propto\lambda^{-1/3}$, which is in contrast to the $T_{c}^{(0)}\propto \lambda^{-1/2}$ for a nonrotating Bose gas. In addition, $T_{c}\propto\Omega^{1/3}$. Let us remind that the critical temperature of a BEC transition for a noninteracting Bose gas in nonrelativistic and ultrarelativistic limits are $T_{c}\propto \Omega^{2/5}$ and $T_{c}\propto \Omega^{1/4}$, respectively \cite{siri2024b}. This demonstrates the effect of rotation in changing the critical exponents of different quantities in the symmetry-broken phase.\\
We defined a reduced temperature $t=T/T_{c}$, and showed that in the symmetry-broken phase, the minimum mentioned above depends on $(1-t^{3})$, while for a nonrotating Bose gas this dependence is $(1-t_{0}^{2})$, where $t_{0}=T/T_{c}^{(0)}$. In the symmetry-restored phase, this minimum vanishes. This indicates a continuous phase transition in both nonrotating and rotating Bose gases. Plugging these minima into $m_{1}^{2}(v)$ and $m_{2}^{2}(v)$, it turned out that at $t\geq 1$, i.e., in the symmetry-restored phase $m_{1}$ and $m_{2}$ are imaginary. Since, according to our arguments in Sec. \ref{sec3}, $m_{2}$ is the mass of a Goldstone mode, we expect that in the chiral limit, i.e., when $m_{0}=0$, it vanishes in the symmetry-broken phase at $t<1$. However, as it is shown in \eqref{E13}, $m_{2}^{2}<0$ in this phase.
\par
To resolve this issue, we followed the method used in \cite{kapusta-book} and added the thermal part of one-loop self-energy diagram to the above results. In contrast to the case of nonrotating bosons, where the thermal mass square is proportional to $\lambda T^{2}$, for rotating bosons it is proportional to $\lambda T^{3}/\Omega$. To arrive at this result, a summation over $\ell$ was necessary. This was performed by utilizing a method originally introduced in \cite{siri2024b}. Adding this perturbative contribution to $m_{i}^{2}, i=1,2$ at $t<1$ and $t\geq 1$, we showed that the Goldstone theorem is satisfied in the chiral limit [see Sec. \ref{sec3C}].
\par
In Secs. \ref{sec3D} and \ref{sec3E}, we added the vacuum and nonperturbative ring potentials to the classical and thermal potentials. The main novelty of these sections lies in the final results for these two parts of the total potential, specifically the method we employed to sum over $\ell$.  According to this method the vacuum part of the potential for a rigidly rotating Bose gas is the same as that for a nonrotating gas. We followed the method described in \cite{carrington1992} to dimensionally regularize the vacuum potential. As concerns the ring potential, we present a novel method to compute this nonperturbative contribution to the thermodynamic potential. In particular, we summed over $\ell$ by performing a $\zeta$-function regularization. In Sec. \ref{sec3F}, we presented a summary of these results.
\par
In Sec. \ref{sec4}, we used the total thermodynamic potential presented in Sec. \ref{sec3} to study the effect of rotation on the spontaneous U(1) symmetry breaking of a realistic model including $\sigma$ and $\pi$ mesons. Fixing free parameters $m_{\sigma},m_{\pi}$, and $\lambda$, and identifying $m_{1}$ and $m_{2}$ with the meson masses $m_{\sigma}$ and $m_{\pi}$, we obtained numerical values for $c$ and $a$ (see Sec. \ref{sec3A}). First, we determined the $T$ dependence of the minima of the total thermodynamic potential, excluding the ring contribution. According to the results presented in Fig. \ref{fig8}, rotation decreases the critical temperature of the U(1) phase transition. Additionally, it is shown that $T_{c}$ increases with increasing $\Omega$.
In \cite{siri2024b}, it is shown that the critical temperature of the BEC in a noninteracting Bose gas under rotation behaves in the same manner. This phenomenon indicates that rotation enhances the condensation.
Recently, a similar result was observed in \cite{wilczek2025}, where it is demonstrated that the interplay between rotation and magnetic fields significantly increases the critical temperature of the superconducting phase transition.
\par
To explore the effect of nonperturbative ring potential, we numerically solved the gap equation corresponding to the  total thermodynamic potential and determined its minima $\bar{v}_{\text{min}}$. Because of the specific form of the ring potential, there was a certain $\bar{v}_{\star}$ through which all the curves $\bar{v}_{\text{min}}(T,\Omega_{\text{f}})$, independent of the chosen $\Omega_{\text{f}}$, converge (see Fig. \ref{fig9}). Moreover, the transition for $\Omega=0$ turned out to be discontinuous, while it is continuous for all $\Omega\neq 0$. As it is demonstrated in Fig. \eqref{fig10}, $T_{\star}$ increases with increasing $\Omega$.
\par
Finally, we determined the $T$ dependence of the masses $m_{\sigma}$ and $m_{\pi}$ mesons for a fixed value of $\Omega$. To achieve this, we utilized \eqref{D7} along with $\bar{v}_{\text{min}}$, which is derived from Figs. \ref{fig8} and \ref{fig9}. The plot shown in Fig. \ref{fig11}(a), based on the total potential excluding the ring contribution, is representative of the $T$ dependence of $m_{\sigma}$ and $m_{\pi}$ (see e.g. \cite{buballa2012}).  However, when we include the ring contribution, the shape of the plots changes, especially at $T > T_{\star}$. The reason is that considering the ring potential changes the order of the phase transition from a second order transition to continuous (for $\Omega\neq 0$) or discontinuous (for $\Omega=0$) a crossover. In this context, we numerically determined the $\sigma$ dissociation temperature $T_{\text{diss}}$, which may serve as an indicator for type of the transition into the symmetry-restored phase. We showed that $T_{\text{diss}}<T_{c}$ and $T_{\text{diss}}^{\star}<T_{\star}$, as expected from a crossover transition \cite{buballa2012}.
\par
It would be intriguing to extend the above findings, in particular those from Sec. \ref{sec3}, to the case of nonvanishing chemical potential. In \cite{alford2008}, the kaon condensation in a certain color-flavor locked phase (CFL) of quark matter is studied at nonzero temperature. This is a state of matter which is believed to exist in quark matter at large densities and low temperatures. Large densities at which the color superconducting CFL phase is built are expected to exist in the interior of neutron stars. One of the main characteristic of these compact stars, apart from densities, is their large angular velocities. It is not clear how a rigid rotation, like that used in the present paper, may affect the formation of pseudo-Goldstone bosons and the critical temperature of the BE condensation in this nontrivial environment. We postpone the study of this problem to our future publication.
\section{Acknowledgments}
The authors thank the organizers of 9th Iranian Conference of Mathematical Physics, where the preliminary results of this paper are presented. They also thank H. Namvar for drawing Fig. \ref{fig6}.

\begin{appendix}
\section{High-temperature expansion of thermodynamic potential}\label{appA}
\setcounter{equation}{0}
In this appendix, we present the high-temperature expansion of following potential
\begin{eqnarray}\label{appA1}
V_{T}=T\sum_{\ell=1}^{\infty}\int d\tilde{k}\ln\left(1-e^{-\beta(\omega+\ell\Omega)}\right),
\end{eqnarray}
with $\omega^{2}\equiv \bs{k}_{\perp}^{2}+k_{z}^{2}+m^{2}$, $\int d\tilde{k}$ defined in \eqref{N20}, and $\Omega>0$. The resulting expressions are then used to evaluate $\mathcal{V}_{T}$ from \eqref{N37}. To begin, we use
\begin{eqnarray}\label{appA2}
\ln(1-x)=-\sum_{j=1}^{\infty}\frac{x^{j}}{j}, \quad\mbox{for $x<1$},
\end{eqnarray}
and rewrite \eqref{appA1} by choosing $x=e^{-\beta(\omega+\ell\Omega)}$, as
\begin{eqnarray}\label{appA3}
V_{T}=-\sum_{j=1}^{\infty}\frac{1}{j}\sum_{\ell=1}^{\infty}e^{-\beta \ell\Omega j}\int d\tilde{k}~e^{-\beta j\omega}.
\end{eqnarray}
The summation over $\ell$ can be carried out by making use of the method first introduced in \cite{siri2024b}. For $\beta j\ell\Omega>0$, the summation over $\ell$ yields
\begin{eqnarray}\label{appA4}
\sum_{\ell=1}^{\infty}e^{-\beta \ell\Omega j}=\frac{1}{1-e^{-\beta \Omega j}}.
\end{eqnarray}
In a slowly rotating Bose gas with $\beta\Omega\ll 1$, we use
\begin{eqnarray}\label{appA5}
\frac{1}{1-e^{-x}}\stackrel{x\ll 1}{\longrightarrow}\frac{1}{x},\quad\mbox{for $x>0$},
\end{eqnarray}
to write
\begin{eqnarray}\label{appA6}
V_{T}=-\frac{T}{\beta\Omega}\sum_{j=1}^{\infty}\frac{1}{j^{2}}\int d\tilde{k}\ e^{-\beta \omega j}.
\end{eqnarray}
Following the method presented in \cite{siri2024a, siri2024b}, we perform the integration over $k$ by replacing $e^{-\beta j\omega}$ with
\begin{eqnarray}\label{appA7}
e^{-\beta \omega j }=\frac{1}{2\pi i}\int_{c-i\infty}^{c+i\infty}dz~\Gamma(z)(\beta j)^{-z}(\omega^{2})^{-z/2},\nonumber\\
\end{eqnarray}
and $(\omega^{2})^{-z/2}$ with
\begin{eqnarray}\label{appA8}
(\omega^{2})^{-z/2}=\frac{1}{\Gamma(z/2)}\int_{0}^{\infty}dt~t^{z/2-1}e^{-\omega^{2} t}.
\end{eqnarray}
Plugging \eqref{appA7} and \eqref{appA8} into \eqref{appA6}, and using
\begin{eqnarray}\label{appA9}
\int\frac{k_{\perp}dk_{\perp}dk_{z}}{(2\pi)^{2}}e^{-(\bs{k}_{\perp}^{2}+k_{z}^{2})t}=\frac{t^{-3/2}}{8\pi^{3/2}},
\end{eqnarray}
and the Legendre formula
\begin{eqnarray}\label{appA10}
\Gamma(z)=\frac{2^{z}}{2\pi^{1/2}}\Gamma\left(\frac{z}{2}\right)\Gamma\left(\frac{z+1}{2}\right),
\end{eqnarray}
we arrive first at
\begin{eqnarray}\label{appA11}
V_{T}&=&-\frac{m^{3}T^{2}}{16\pi^{2}\Omega}\frac{1}{2\pi i}\int_{c-i\infty}^{c+i\infty}dz\ \zeta(2+z)\left(\frac{\beta m}{2}\right)^{-z}\nonumber\\
&&\times\Gamma\left(\frac{z+1}{2}\right)\Gamma\left(\frac{z-3}{2}\right).
\end{eqnarray}
Here, $\sum\limits_{j=1}^{\infty}j^{-(2+z)}=\zeta(2+z)$ with $\zeta(z)$ the Riemann $\zeta$-function, and
\begin{eqnarray}\label{appA12}
\int_{0}^{\infty}dt\ t^{-5/2+z/2}  e^{-m^{2} t}=m^{3-z}\Gamma\left(\frac{z-3}{2}\right),
\end{eqnarray}
are used. Finally, the Mellin-Barnes integral over $z$ in \eqref{appA11} yields
\begin{eqnarray}\label{appA13}
V_{T}&=&-\frac{T^{5}\zeta(5)}{\pi^{2}\Omega}+\frac{T^{3}m^{2}\zeta(3)}{4\pi^{2}\Omega}-\frac{T m^{4}}{384\Omega}-\frac{7T m^{4}}{256\pi^{2}\Omega}\nonumber\\
&&+\frac{T m^{4}\gamma_{E}^{2}}{32\pi^{2}\Omega}+\frac{T m^{4}\gamma_{1}}{16\pi^{2}\Omega}
+\frac{3T m^{4}}{64\pi^{2}\Omega}\ln\left(\frac{m\beta}{2}\right)\nonumber\\
&&-\frac{T m^{4}}{32\pi^{2}\Omega}\left(\ln\left(\frac{m\beta}{2}\right)\right)^{2}+\cdots,
\end{eqnarray}
where $\gamma_{1}$ is the coefficient of $(s-1)$ in the Laurent expansion of $\zeta(s)$ about the point $s=1$,
\begin{eqnarray}\label{appA14}
\hspace{-0.5cm}
\zeta(s)=\frac{1}{s-1}+\gamma_{E} -(s-1) \gamma _1+\mathcal{O}\left((s-1)^2\right).
\end{eqnarray}
In Sec. \eqref{sec3}, the first two terms of the high-temperature expansion of $V_{T}$ from \eqref{appA13} are used to study the spontaneous breaking of global $\mathrm{U}(1)$ symmetry in $\lambda(\varphi^{\star}\varphi)$ model.
\section{Derivation of \eqref{E27}}\label{appB}
\setcounter{equation}{0}
In Sec. \eqref{sec3}, we arrived at $\Pi_{i}^{\text{mat}}$ from \eqref{E26},
\begin{eqnarray}\label{appB1}
\Pi_{i}^{\text{mat}}=\frac{\lambda}{\beta\Omega}\sum_{j=1}^{\infty}\frac{I_{ij}}{j},
\end{eqnarray}
with
\begin{eqnarray}\label{appB2}
I_{ij}\equiv \int d\tilde{k}~\frac{e^{-\beta \omega_{i}j}}{\omega_{i}},
\end{eqnarray}
and $\omega_{i}^{2}=\bs{k}_{\perp}^{2}+k_{z}^{2}+m_{i}^{2}$. In this appendix, we derive the final result \eqref{E27} for the one-loop self-energy $\Pi_{i}^{\text{mat}}$. To evaluate the $k$-integration in \eqref{appB2}, we use \eqref{appA7} to arrive first at
\begin{eqnarray}\label{appB3}
I_{ij}=\frac{1}{2\pi i}\int d\tilde{k}\int_{c-i\infty}^{c+i\infty} dz \Gamma(z)\left(\beta j\right)^{-z}\left(\omega_{i}^{2}\right)^{-(z+1)/2}.\nonumber\\
\end{eqnarray}
Replacing $(\omega_{i}^{2})^{-(z+1)/2}$ in \eqref{appB3} with
\begin{eqnarray}\label{appB4}
(\omega_{i}^{2})^{-(z+1)/2}=\frac{1}{\Gamma\left(\frac{z+1}{2}\right)}\int_{0}^{\infty}dt~t^{(z+1)/2-1}e^{-\omega_{i}^{2} t}, \nonumber\\
\end{eqnarray}
and performing the $k$-integration by making use of \eqref{appA9}, $I_{ij}$ is given by
\begin{eqnarray}\label{appB5}
I_{ij}=\frac{m_{i}^{2}}{16\pi^{2}}\frac{1}{2\pi i}\int_{c-i\infty}^{c+i\infty}dz~\Gamma\left(\frac{z}{2}\right)\Gamma\left(\frac{z-2}{2}\right)\left(\frac{\beta m_{i}j}{2}\right)^{-z}. \nonumber\\
\end{eqnarray}
Here, the Legendre formula \eqref{appA10} and
\begin{eqnarray}\label{appB6}
\int_{0}^{\infty}dt~ t^{(z-2)/2-1}e^{-m_{i}^{2}t}=\left(m_{i}^{2}\right)^{-z/2+1}\Gamma\left(\frac{z-2}{2}\right),\nonumber\\
\end{eqnarray}
are utilized. Plugging at this stage \eqref{appB6} into \eqref{appB1} and using
$
\sum\limits_{j=1}^{\infty}j^{-(1+z)}=\zeta(1+z),
$
we obtain
\begin{eqnarray}\label{appB7}
\lefteqn{\Pi_{i}^{\text{mat}}=\frac{\lambda T m_{i}^{2}}{16\pi^{2}\Omega}}\nonumber\\
&&\times
\frac{1}{2\pi i}\int_{c-i\infty}^{c+i\infty}dz~\zeta(1+z)\left(\frac{z}{2}\right)\Gamma\left(\frac{z-2}{2}\right)\left(\frac{\beta m_{i}}{2}\right)^{-z}\nonumber\\
&=&\frac{\lambda T^{3}\zeta(3)}{2\pi^{2}\Omega}+\cdots.
\end{eqnarray}
At high temperatures, the first term in \eqref{appB7} is the most dominant thermal mass correction to $m_{i}^{2}$, as is described in Sec. \ref{sec3B}.
\section{Derivation of \eqref{E34} in cylinder coordinate system}\label{appC}
\setcounter{equation}{0}
In this appendix, we evaluate the integrals of the form
\begin{eqnarray}\label{appC1}
\Phi(m,d,n)=\int \frac{d\tilde{k}}{\left(\boldsymbol{k}_\perp^2+k_z^2+m^2 \right)^{n}},
\end{eqnarray}
in cylindrical coordinate system by an appropriate $d$-dimensional regularization. To this purpose, we replace $d\tilde{k}$ with $\frac{d^{d}{k}}{(2\pi)^d}$, where $d=3-\epsilon$. Here, $\epsilon$ is an infinitesimal regulator.
In cylindrical coordinate the volume element in momentum space $d^{d}{k}$ reads $d^{d}k=d{k_\bot}k_\bot^{d-2}d{\Omega}_{d-1}d{k_z}$, where the $d$-dimensional solid angle $d{\Omega}_{d-1}$ is given by
\begin{eqnarray}\label{appC2}
d{\Omega}_{d-1}\equiv\dfrac{2\pi^{\frac{d-1}{2}}}{\Gamma\left(\frac{d-1}{2}\right)}.
\end{eqnarray}
Using, at this stage, the Schwinger parametrization
\begin{eqnarray}\label{appC3}
\frac{1}{\left( \boldsymbol{k}_\bot^2+k_z^2+m^2 \right)^{n}}=\frac{1}{\Gamma(n)}\int_{0}^{\infty}d{t}~t^{n-1}e^{-t\left(\bs{k}_\bot^2+k_z^2+m^2\right)},\nonumber\\
\end{eqnarray}
we can write \eqref{appC1} as
\begin{eqnarray}\label{appC4}
\Phi(m,d,n)&=&\dfrac{2\pi^{(d-1)/2}}{(2\pi)^d\Gamma\left(\frac{d-1}{2}\right)\Gamma(n)}\int_{0}^{\infty}d{k_\bot}k_\bot^{d-2}\int_{-\infty}^{+\infty}d{k_z}\nonumber\\
&&\times\int_{0}^{\infty}d{t}~t^{n-1}e^{-t\left(\bs{k}_\bot^2+k_z^2+m^2\right)}.
\end{eqnarray}
To perform the integration over $k_{z}$ and $k_{\perp}$, we use following Gaussian integrals:
\begin{eqnarray}\label{appC5}
\int_{-\infty}^{+\infty}d{k_z}e^{-t k_z^{2}}&=&\left(\dfrac{\pi}{t}\right)^{1/2},\nonumber\\
\int_{0}^{\infty}d{k_\bot}k_\bot^{d-2}e^{-t \bs{k}_\bot^{2}}&=&\frac{t^{(d-1)/2}}{2}\Gamma\left(\frac{d-1}{2}\right).
\end{eqnarray}
By substituting these results into \eqref{appC4}, we arrive at \eqref{E34},
\begin{eqnarray}\label{appC6}
\Phi(m,d,n)=\frac{1}{(4\pi)^{d/2}}\frac{\Gamma\left(n-d/2\right)}{\Gamma(n)}\left(m^2\right)^{-n+d/2}.
\end{eqnarray}
\section{Derivation of \eqref{E44}}\label{appD}
\setcounter{equation}{0}
In this appendix, we outline the derivation of \eqref{E44}. In particular, we focus on the combinatorial factors. Let us start with $\mathcal{V}_{ring}^{A}$. According to its definition, there are $N$ insertions of $\Pi_{2}$ and $N$ propagators $D_{1}$\footnote{Here, the notation $D_{i}\equiv D_{\ell}(\omega_{n},\omega_{i})$ is used.} (see Fig. \ref{fig6}).  Having in mind that for a vertex of type 3 in Fig. \ref{fig3}, each factor $\frac{\lambda}{2}\times 2$ belongs to a $\Pi_{2}$ insertion in a ring with $D_{1}$ propagator, we obtain
\begin{eqnarray}\label{appD1}
\mbox{Type A:}\quad\left(-\frac{\lambda}{2}\times 2\right)^{N}\frac{(N-1)!}{2 N!}\to \frac{(-\Pi_{2})^{N}}{2N}.
\end{eqnarray}
The ring potential of type A is thus given by
\begin{eqnarray}\label{appD2}
\mathcal{V}_{\text{ring}}^{\text{A}}=-\frac{T}{2}\sum_{n,\ell}\int d\tilde{k}\sum_{N=2}^{\infty}\frac{1}{N}\left(-\Pi_{2}D_{1}\right)^{N}.
\end{eqnarray}
Similarly, the combinatorial factor of $\mathcal{V}_{\text{ring}}^{B}$ from Fig. \ref{fig6}, including $N$ insertions of $\Pi_{1}$ and $N$ propagators $D_{1}$ is given by \eqref{appD1} with $\Pi_{2}$ replaced with $\Pi_{1}$
\begin{eqnarray}\label{appD3}
\mbox{Type B:}\quad\left(-\frac{\lambda}{2}\times 2\right)^{N}\frac{(N-1)!}{2 N!}\to \frac{(-\Pi_{1})^{N}}{2N}.
\end{eqnarray}
Here, similar to the previous case, for a vertex of type 3 in Fig. \ref{fig3}, each factor $\frac{\lambda}{2}\times 2$ belongs to a $\Pi_{1}$ insertion in a ring with $D_{2}$ propagator.
For the ring potential of type B, we thus obtain
\begin{eqnarray}\label{appD4}
\mathcal{V}_{\text{ring}}^{\text{B}}=-\frac{T}{2}\sum_{n,\ell}\int d\tilde{k}\sum_{N=2}^{\infty}\frac{1}{N}\left(-\Pi_{1}D_{2}\right)^{N}.
\end{eqnarray}
As concerns the ring potential of type C, which is defined by $r$ insertions of $\Pi_{2}$ and $s$ insertions of $\Pi_{1}$ with $N$ propagators $D_{2}$. Here, $r\geq 1$ and $r+s=N$. For the corresponding combinatorial factor, we arrive first at
\begin{eqnarray}\label{appD5}
\lefteqn{\mbox{Type C}:}\nonumber\\
&\quad&\left(-\frac{\lambda}{4}\times 3!\times 2\right)^{r}\left(-\frac{\lambda}{2}\times 2\right)^{N-r}~ \frac{(N-r)! (r-1)!}{2N!}\nonumber\\
&&\to \left(-\Pi_{2}\right)^{r}\left(-\Pi_{1}\right)^{N-r}~\frac{(N-r)! (r-1)!}{2 N!}.
\end{eqnarray}
Here, the factor $3!\times 2$ is the corresponding combinatorial factor to $\Pi_{2}$ inserted in a ring with $D_{2}$ propagator. For the ring of type $C$, we get
\begin{eqnarray}\label{appD6}
\mathcal{V}_{\text{ring}}^{\text{C}}&=&-\frac{T}{2}\sum_{n,\ell}\int d\tilde{k}\sum_{N=2}^{\infty}\sum_{r=1}^{N}\frac{(N-r)! (r-1)!}{N!}\nonumber\\
&&\times \big[\left(-\Pi_{2}\right)^{r}\left(-\Pi_{1}\right)^{N-r}D_{2}^{N}\big].
\end{eqnarray}
Similar arguments for $\mathcal{V}_{\text{ring}}^{D}$ with $r$ insertions of $\Pi_{1}$, $s$ insertions of $\Pi_{2}$ and $N$ propagators $D_{1}$ lead first to
\begin{eqnarray}\label{appD7}
\lefteqn{\mbox{Type D}:}\nonumber\\
&\quad&\left(-\frac{\lambda}{4}\times 3!\times 2\right)^{r}\left(-\frac{\lambda}{2}\times 2\right)^{N-r}~ \frac{(N-r)! (r-1)!}{2N!}\nonumber\\
&&\to \left(-\Pi_{1}\right)^{r}\left(-\Pi_{2}\right)^{N-r}~\frac{(N-r)! (r-1)!}{2 N!},
\end{eqnarray}
and then to
\begin{eqnarray}\label{appD8}
\mathcal{V}_{\text{ring}}^{\text{D}}&=&-\frac{T}{2}\sum_{n,\ell}\int d\tilde{k}\sum_{N=2}^{\infty}\sum_{r=1}^{N}\frac{(N-r)! (r-1)!}{ N!}\nonumber\\
&&\times \big[\left(-\Pi_{1}\right)^{r}\left(-\Pi_{2}\right)^{N-r}D_{1}^{N}\big].
\end{eqnarray}
\end{appendix}

\end{document}